  \documentclass{EngC}

  \usepackage[rightcaption,raggedright]{sidecap}
  \usepackage{framed}         
  \usepackage{soul}           

  \usepackage{rotating}
  \usepackage{floatpag}
  \rotfloatpagestyle{empty}

 \usepackage{amsmath}
  \usepackage{amsthm}
  \usepackage{graphicx} 

  \usepackage{multind}
  \makeindex{authors}
  \makeindex{subject}

  \theoremstyle{plain}

  \theoremstyle{definition}

  \theoremstyle{remark}

\usepackage{acronym}
\usepackage[ruled,linesnumbered]{algorithm2e}  
\SetKwInput{KwData}{\textbf{Init}} 
\usepackage{amssymb}

\newcommand{\myVec}[1]{{\boldsymbol{#1}}}
\newcommand{\myMat}[1]{{\mathsf{#1}}}
\newcommand{\mySet}[1]{\mathcal{#1}}
\newcommand{\myI}{{\myMat{I}}}			 		
	
\newcommand{\myY}{{\myVec{Y}}}			 		
\newcommand{\myS}{{\myVec{S}}}			 		
\newcommand{\myState}{\bar{\myS}} 
\newcommand{\myStateR}{\bar{\myVec{s}}}
\newcommand{\FwdMsg}[2]{\overrightarrow{\mu}_{#2}}
\newcommand{\BwdMsg}[2]{\overleftarrow{\mu}_{#2}}

\newcommand{\Pdf}[1]{p_{ { #1}} }

\newcommand{\Mem}{L}			 			
\newcommand{\BRNNLen}{B}
\newcommand{\Blklen}{T}			 			
\newcommand{\Blkset}{\mySet{T}}
\newcommand{\CnstSize}{M}			 			

\newcommand{\CovMat}[1]{\myMat{\Sigma}_{#1}}			

\newcommand{\Vecdim}[1]{} 

\newcommand{\eqspace}{\vspace{-0.2cm}}

\newcommand{\Nusers}{K}
\newcommand{\Nantennas}{N}
\newcommand{\NusersSet}{\mySet{K}}

\newcommand{\SigW}{\sigma_w^2}
\newcommand{\SigE}{\sigma_e^2}
\newcommand{\Niter}{Q}
\newcommand{\NiterSet}{\mySet{Q}}
\newcommand{\Ntraining}{n_t}

\newcommand{\includefig}[1]{\includegraphics[width = \textwidth]{#1}}

\acrodef{adc}[ADC]{analog-to-digital convertor}
\acrodef{cs}[CS]{compressed sensing}
\acrodef{dtft}[DTFT]{discrete-time Fourier transform}
\acrodef{dnn}[DNN]{deep neural network} 
\acrodef{csi}[CSI]{channel state information}
\acrodef{map}[MAP]{maximum a-posteriori probability}
\acrodef{snr}[SNR]{signal-to-noise ratio}
\acrodef{bs}[BS]{base station} 
\acrodef{iot}[IOT]{Interent of Things}
\acrodef{mimo}[MIMO]{multiple-input multiple-output}
\acrodef{mse}[MSE]{mean-squared error}
\acrodef{mmse}[MMSE]{minimum \ac{mse}}
\acrodef{pdf}[PDF]{probability density function}
\acrodef{pmf}[PMF]{probability mass function}
\acrodef{rv}[RV]{random variable}
\acrodef{fec}[FEC]{forward error correction}
\acrodef{rs}[RS]{Reed-Solomon}
\acrodef{lti}[LTI]{linear time-invariant}
\acrodef{wss}[WSS]{wide-sense stationary}
\acrodef{psd}[PSD]{power spectral density}
\acrodef{ser}[SER]{symbol error rate} 
\acrodef{ber}[BER]{bit error rate} 
\acrodef{sgd}[SGD]{stochastic gradient descent} 
\acrodef{isi}[ISI]{intersymbol interference}  
\acrodef{awgn}[AWGN]{additive white Gaussian noise} 
\acrodef{ut}[UT]{user terminal} 
\acrodef{mmw}[mmWave]{millimeter wave}
\acrodef{cfl}[CFL]{clustered \ac{fl}} 
\acrodef{fl}[FL]{federated learning}
\acrodef{pic}[PIC]{principal inertia component}
\acrodef{ml}[ML]{machine learning}
\acrodef{lstm}[LSTM]{long short-term memory} 
\acrodef{em}[EM]{expectation minimization} 
\acrodef{awgn}[AWGN]{additive white Gaussian noise}
\acrodef{rnn}[RNN]{recurrent neural network}
\acrodef{sbrnn}[SBRNN]{sliding bidirectional \ac{rnn}}
\acrodef{sic}[SIC]{soft interference cancellation}
\acrodef{hmm}[HMM]{hidden Markov model}
\acrodef{bpsk}[BPSK]{binary phase shift keying}
\acrodef{gru}[GRU]{gated recurrent unit}

\definecolor{NewColor}{rgb}{0,0,0}

\def\kth{$k$th}
\def\ith{$i$th}	
\def\jth{$j$th}

\usepackage{lipsum}

\begin{document}



  \mainmatter 



\alphafootnotes
\author[N\, Shlezinger, N\, Farsad, Y\,C Eldar, and A\,J Goldsmith]
{Nir Shlezinger, Nariman Farsad, Yonina C. Eldar, and Andrea J. Goldsmith}

\chapterauthor{Nir Shlezinger,
	Nariman Farsad,
	Yonina C. Eldar, and
	Andrea J. Goldsmith}

\chapter{Model-Based Machine Learning for Communications}


\contributor{Nir Shlezinger
	\affiliation{Ben-Gurion University,
		School of Electrical and Computer Engineering,
		Be'er-Sheva, Israel}}

\contributor{Nariman Farsad
	\affiliation{Ryerson University,
		Department of Computer Science,
		Toronto, Canada}}

\contributor{Yonina C. Eldar
	\affiliation{Weizmann Institute of Science,
		Faculty of Mathematics and Computer Science,
		Rehovot, Israel}}

\contributor{Andrea J. Goldsmith
	\affiliation{Stanford University,
		Department of Electrical Engineering,
		Palo Alto, CA}}

\section{Introduction}
Traditional communication systems design is dominated by methods that are based on statistical models. These statistical-model-based algorithms, which we refer to henceforth as {\em model-based methods}, rely on mathematical models that describe the transmission process, signal propagation, receiver noise, interference, and many other components of the system that affect the end-to-end signal transmission and reception. Such mathematical models use parameters that vary over time as the channel conditions, the environment, network traffic, or network topology change. Therefore, for optimal operation, many of the algorithms used in communication systems rely on the underlying mathematical models as well as the estimation of the model parameters.  However, there are cases where this approach fails, in particular when the mathematical models for one or more of the system components are highly complex, hard to estimate, poorly understood, do not well-capture the underlying physics of the system, or do not lend themselves to computationally-efficient algorithms.   
In some other cases, although mathematical models are known, accurate parameter estimation may not be possible.    
Finally, common hardware limitations, such as the restriction to utilize low-resolution quantizers or non-linear power amplifiers, can significantly increase the complexity of the underlying channel model.  

An alternative data-driven approach is based on \ac{ml}. \ac{ml} techniques, and in particular, deep learning, have been the focus of extensive research in recent years due to their empirical success in various applications, including computer vision and speech processing \cite{lecun2015deep, hinton2012deep}. 
The benefits of \ac{ml}-driven methods over traditional model-based approaches are threefold: 
First, \ac{ml} methods are independent of the underlying stochastic model, and thus can operate efficiently in scenarios where this model is unknown or its parameters cannot be accurately estimated.
Second, when the underlying model is extremely complex, \ac{ml} algorithms have demonstrated the ability to extract and disentangle the meaningful semantic information from the observed data \cite{Bengio09learning}, a task which is very difficult to carry out using traditional model-based approaches, even when the model is perfectly known. 
Finally, the main complexity in utilizing \ac{ml} methods is in the training stage, which is typically carried out offline. Once trained, they tend to implement inference at a  lower computational burden and delay compared to their analytical model-based counterparts \cite{gregor2010learning}.

Although \ac{ml} has been the focus of significant research attention over the last decade, it has yet to significantly contribute to practical designs in one of the most important technologies of the modern era -- digital communication. 
The fact that \ac{ml}-based algorithms, which have revolutionized the fields of computer vision and natural language processing, do not yet play a fundamental role in the design of physical layer communication systems, and particularly digital receivers, may be due to one or more of the following reasons: 
\begin{enumerate}
	\item The large amount of possible outputs impose a major challenge in efficiently applying \ac{ml} algorithms. In particular, the constellation size of the modulation and the blocklength of the channel code, combined with the time-varying nature of communication channels, leads to an exponentially large number of possible channel outputs that an \ac{ml} receiver algorithm must be trained on. 
	\item {Traditional deep learning techniques require high computational resources, while communication devices, such as wearable devices and mobile phones, are typically limited in hardware and power.}
	\item To date, conventional communication schemes, which assume a simplified channel model with parameters that are dynamically estimated, have been very successful.
\end{enumerate}
The third reason is likely to become less relevant as the spectrum congestion of existing cellular standards forces future communication systems to explore new frequency ranges and share spectrum with other application such as radar \cite{zheng2019radar}.  As these new frequency bands and spectrum sharing techniques become widespread, the simplified channel, interference, and noise models used in current communication receiver techniques may no longer work well. Moreover, the strict cost, power, and memory constraints imposed on communicating devices lead to the usage of low-resolution \acp{adc} and power amplifiers with dominant non-linearities \cite{singya17PA}. This makes the successful application of model-based techniques significantly more complex.  Thus, conventional model-based approaches may no longer be able to meet the performance and throughput demands of future wireless devices, motivating their combination with data-driven approaches based on \ac{ml} such as deep learning. Such techniques must still overcome the challenges identified above with respect to the large computational resources and data sets needed for training.

\begin{figure}
	\centering
	\includefig{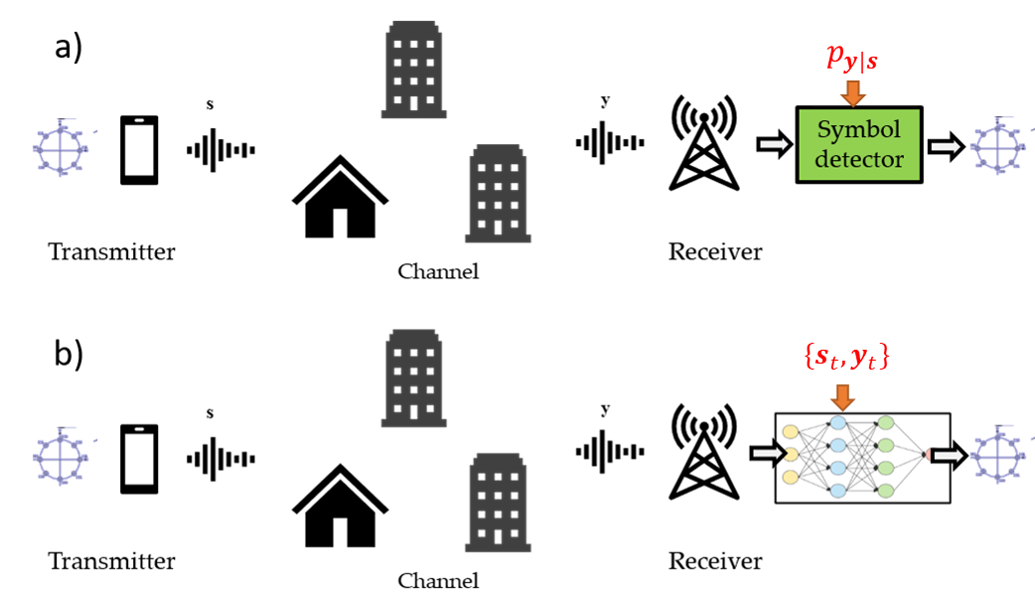} 
	\caption{Model-based methods versus deep learning for symbol detection: $a)$ a receiver uses its knowledge of the underlying statistical channel model, denoted $\Pdf{\myY|\myS}$, to detect the transmitted symbols for the channel output in a model-based manner; $b)$ a receiver uses a \ac{dnn} trained using the data set $\{\myVec{s}_t, \myVec{y}_t\}$ for recovering the symbols.} 
	\label{fig:SymDetComp1}
\end{figure}

Despite its unprecedented success, deep learning is subject to several challenges which limit its applicability in some important communication scenarios. In particular, \acp{dnn} consist of highly-parameterized systems that can represent a broad range of mappings. As such, massive data sets are typically required to learn a desirable mapping, and the computational burden of training and utilizing these networks may constitute a major drawback. For example, consider the two receivers illustrated in Fig. \ref{fig:SymDetComp1}, which carry out symbol detection using model-based  algorithms and model-agnostic \acp{dnn}, respectively. The dynamic nature of wireless channels implies that the receivers should track channel variations in order to  reliably detect the transmitted messages over long periods of time. To do so, the model-based receiver in Fig. \ref{fig:SymDetComp1}(a) typically estimates the model parameters imposed on the underlying statistics using periodic pilots. For the same purpose, the \ac{dnn}-based receiver in Fig. \ref{fig:SymDetComp1}(b)  should periodically re-train its \ac{dnn} to track channel variations. The fact that doing so requires a large data set  leads to a significant decrease of spectral efficiency and increase in computational complexity associated with this  training. Furthermore, \acp{dnn} are commonly utilized as black-boxes, and thus do not  offer the interpretability, flexibility, versatility, and reliability of model-based techniques.

The limitations associated with model-based methods and black-box deep learning systems gave rise to a set of techniques on the interface of traditional model-based communication and \ac{ml}, attempting to benefit from the best of both worlds \cite{shlezinger2020model}. Such {\em model-based \ac{ml}} systems can be divided into two main categories. The first of the two  utilizes model-based methods as a form of domain knowledge in designing a \ac{dnn} architecture, which is then trained and used for inference. The most common example of this strategy is the family of {\em deep unfolded networks} \cite{hershey2014deep, monga2019algorithm}, which design the layers of a \ac{dnn} to imitate the iterations of a model-based optimization algorithm, and has been utilized in various communication-related tasks \cite{balatsoukas2019deep}.  
The second strategy, which we call {\em \ac{dnn}-aided hybrid algorithms}, uses model-based methods with integrated \acp{dnn} for inference by incorporating \ac{ml} in a manner that makes the system more robust and model-agnostic. Communication receivers designed using \ac{dnn}-aided inference include   the data-driven implementations of the Viterbi algorithm \cite{shlezinger2019viterbinet} and the BCJR detector \cite{shlezinger2020data}. 

In this chapter we present an introduction to model-based \ac{ml} for communication systems. We begin by reviewing existing strategies for combining model-based algorithms and \ac{ml} from a high level perspective in Section~\ref{sec:ModelML}, and compare them to the conventional deep learning approach which utilizes established \ac{dnn} architectures trained in an end-to-end manner. Then, in Section~\ref{sec:SymbolDetection} we focus on symbol detection, which is one of the fundamental tasks of communication receivers. We show how each strategy, i.e., conventional \ac{dnn} architectures, deep unfolding, and \ac{dnn}-aided hybrid algorithms, can be applied to this problem. The last two approaches constitute a middle ground between the purely model-based and the \ac{dnn}-based receivers illustrated in Fig. \ref{fig:SymDetComp1}. By focusing on this specific task, we highlight the advantages and drawbacks of each strategy, and present guidelines to facilitate the design of future model-based deep learning systems for communications. We conclude this chapter with a summary provided in Section~\ref{sec:Summary}.

\section{Model-Based Machine Learning}
\label{sec:ModelML}
We begin by reviewing the leading approaches for combining \ac{ml}, and particularly deep learning, with model-based algorithms. 
The neural networks in this hybrid-approach are trained in a supervised manner and then used during inference. Then, in Section \ref{sec:SymbolDetection} we focus on symbol detection and provide concrete examples on how this approach can be used to design data-driven detectors. 

In a broad family of problems, a system is required to map an input variable $\myVec{Y}$  into a prediction of a label variable $\myVec{S}$. \ac{ml} systems learn such a mapping from a training set consisting of $\Ntraining$ pairs of inputs and their corresponding labels, denoted $\{\myVec{s}_t, \myVec{y}_t\}_{t=1}^{\Ntraining}$. Model-based methods carry out such  inference based on prior knowledge of the statistical model relating $\myVec{Y}$ and $\myVec{S}$, denoted $\Pdf{\myY|\myS}$. Model-based \ac{ml} systems reviewed in this chapter combine model-based methods with learning techniques, namely, they tune their mapping of the input $\myVec{Y}$ based on both a labeled training set  $\{\myVec{s}_t, \myVec{y}_t\}_{t=1}^{\Ntraining}$ as well as some knowledge of the underlying distribution. Such hybrid data-driven model-aware systems can typically learn their mappings from smaller training sets compared to purely model-agnostic \acp{dnn}, and commonly operate without full and accurate knowledge of  $\Pdf{\myY|\myS}$, upon which model-based methods are based. We next elaborate on the main strategies of combining \ac{ml} and model-based techniques, beginning with extreme cases of \acp{dnn} that rely solely on data and purely model-based inference algorithms.  

\subsection{Conventional Deep Learning}
\label{subsec:ModelMLDeep}
The conventional application of deep learning is to carry out inference using some standard \ac{dnn} architecture. This \ac{dnn} uses the training data to learn how to map a realization of the input  $\myVec{Y}= \myVec{y}$ into a prediction $\hat{\myVec{s}}$. Such highly-parameterized networks can effectively approximate any Borel measurable mapping, as it follows from the universal approximation theorem \cite[Ch. 6.4.1]{goodfellow2016deep}. Therefore, by properly tuning their parameters using a {\em sufficiently large} training set, typically using optimization based on some variant of \ac{sgd}, one should be able to obtain the desirable inference rule. 

While standard \acp{dnn} structures are highly model-agnostic and are commonly treated as black-boxes, one can still incorporate some level of domain knowledge in the selection of the specific network architecture. For instance, when the input is known to exhibit temporal correlation, architectures based on \acp{rnn} or transformers are known to be preferable. Alternatively, in the presence of spatial patterns, one may prefer to utilize convolutional layers. An additional method to incorporate domain knowledge into a black-box \ac{dnn} is by pre-processing of the input via, e.g., feature extraction. 

Conventional deep learning based on established black-box \acp{dnn} is data-driven, i.e, it requires  data representing the problem at hand, possibly combined with a very basic level of domain knowledge to select the specific architecture.
A major drawback of using such networks, which is particularly relevant in the context of communication systems, is that learning a large number of parameters requires a massive data set to train. In dynamic environments, even when a sufficiently large data set is available, it is difficult to train a model that performs optimally over the whole range of the dynamically changing system. Moreover, online training as the system dynamics change tends to be computationally expensive because of the large number of parameters.

\subsection{Model-Based Methods}
\label{subsec:ModelMLModel}
\begin{figure}
	\centering
	\includefig{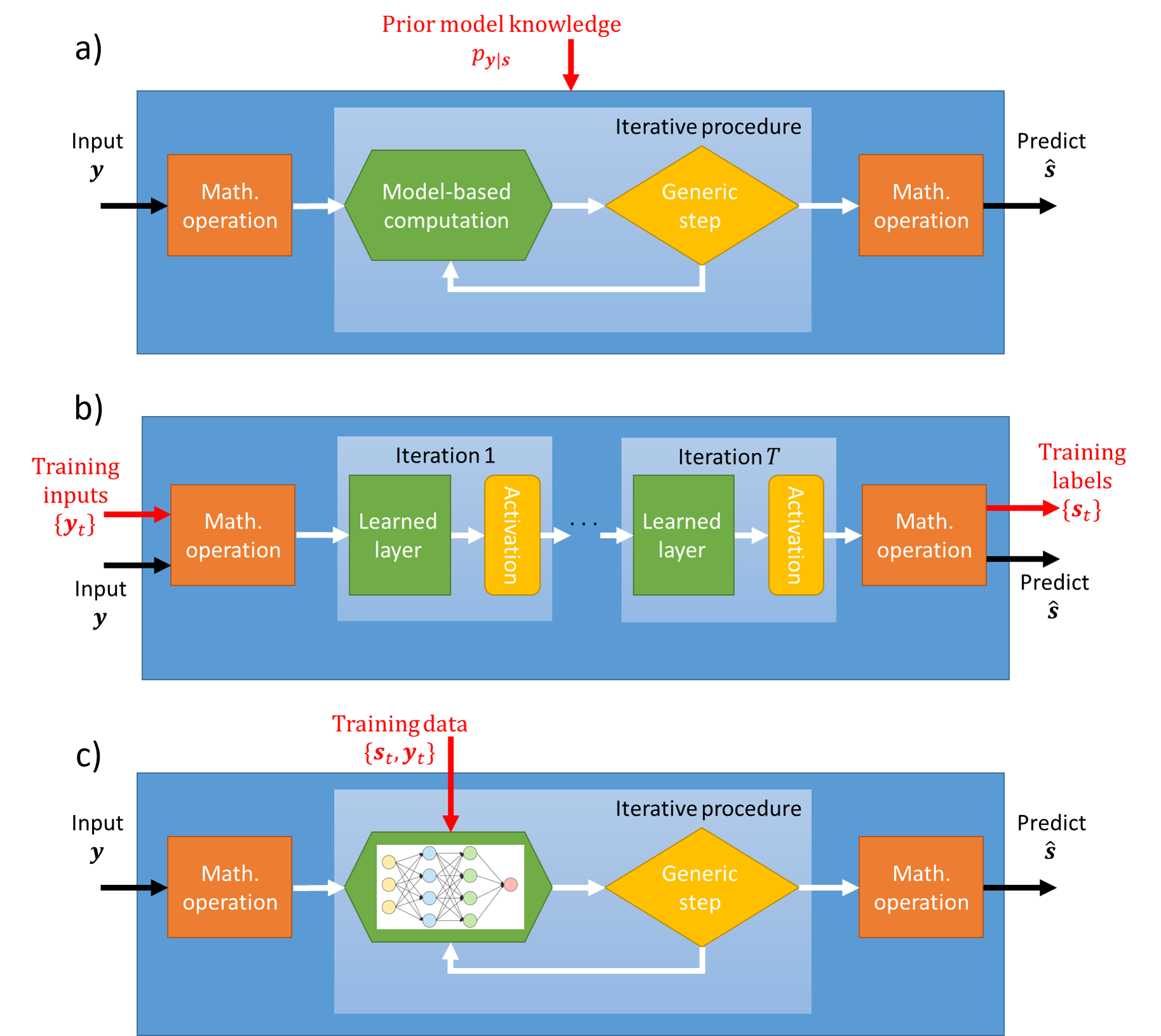} 
	\caption{Illustration of $a)$ model-based iterative algorithm; $b)$ unfolding the algorithm into a \ac{dnn}; $c)$ converting the scheme into a \ac{dnn}-aided method.} 
	\label{fig:ModelBasedFlow}
\end{figure}

Model-based algorithms carry out inference based on prior knowledge of the underlying statistics relating the input $\myVec{Y}$ and the label $\myVec{S}$, i.e., $\Pdf{\myVec{Y} | \myVec{S}}$. A common family of model-based methods is based on iterative algorithms, which allow us to infer with provable performance and controllable complexity in an iterative fashion, as illustrated in Fig. \ref{fig:ModelBasedFlow}(a). This generic family of iterative algorithms  consists of some input and output processing stages, with an intermediate iterative procedure. The latter can in turn be divided into a model-based computation, namely, a procedure that is determined by $\Pdf{\myVec{Y} | \myVec{S}}$; and a set of generic mathematical manipulations. 

These algorithms vary significantly between different statistical models. For instance, for the symbol detection task, model-based methods such as the Viterbi detector \cite{viterbi1967error} or the BCJR algorithm \cite{bahl1974optimal} are valid for finite-memory channels, while for \ac{mimo} detection one may utilize the family of interference cancellation methods  \cite{andrews2005interference}. Each such algorithm may be specifically tailored to a given scenario, as opposed to black-box \acp{dnn}, in which the parameterized inference rule is generic, and the unique characteristics of the scenario at hand are encapsulated in the parameters learned during training.

Model-based techniques do not rely on data to learn their mapping, though data is often used to estimate unknown model parameters. In practice, accurate knowledge of the statistical model relating the observations and the desired information is typically unavailable, and thus applying such techniques commonly requires imposing some assumptions on the underlying statistics, which  in some cases  reflects the actual behavior, but often  do not. In the presence of inaccurate knowledge of $\Pdf{\myVec{Y} | \myVec{S}}$ due to estimation errors or due to enforcing a model that does not fully capture the environment, the performance of model-based methods tends to degrade considerably. This limits the applicability of model-based algorithms in scenarios where  $\Pdf{\myVec{Y} | \myVec{S}}$ is unknown, costly to estimate accurately, or too complex to express analytically.

\subsection{Model-Based Deep Learning by Deep Unfolding} 
\label{subsec:ModelMLUnfolding}
Deep unfolding \cite{hershey2014deep,monga2019algorithm}, also referred to as {\em deep unrolling}, is a common strategy to combine deep learning with model-based algorithms. Here, model-based methods are utilized as a form of domain knowledge in designing a \ac{dnn} architecture, which is trained end-to-end and then used for inference. Unlike the application of conventional black-box \acp{dnn} discussed in Section \ref{subsec:ModelMLDeep}, deep unfolding utilizes a unique \ac{dnn} structure designed specifically for the task at hand. 

The main rationale in deep unfolding is to design the network to imitate the operation of a model-based iterative optimization algorithm corresponding to the considered problem. In particular, each iteration of the model-based algorithm is replaced with a dedicated layer with trainable parameters whose structure is based on the operations carried out during that iteration. An illustration of a neural network obtained by unfolding the model-based iterative model of Fig.~\ref{fig:ModelBasedFlow}(a) is depicted in Fig.~\ref{fig:ModelBasedFlow}(b), where a network with $T$ layers is designed to imitate $T$ iterations of the optimization method. Once the architecture is fixed, the resulting network is trained in an end-to-end manner as in conventional deep learning.

{
Deep unfolded networks, in which the iterations consist of trainable parameters, are typically capable of inferring with a smaller number of layers compared to the amount of iterations required by the model-based algorithm. Consequently,  even when the model-based algorithm is feasible, processing $\myVec{Y}$ through a trained unfolded \ac{dnn} is typically faster than applying the iterative algorithm \cite{gregor2010learning}. Furthermore, converting a model-based algorithm into an unfolded deep network can also improve its performance. For example, iterative algorithms based on some relaxed optimization commonly achieve improved accuracy when unfolded into a \ac{dnn}, due to the ability to learn to overcome the error induced by relaxation in the training stage. 
The main benefits of deep unfolding over using end-to-end networks stem from its incorporation of domain knowledge in the network architecture. As such, unfolded networks can achieve improved performance at reduced complexity, i.e., when operating with less parameters, compared to conventional end-to-end networks \cite{balatsoukas2019deep}.}
Nonetheless, deep unfolded networks are  highly parameterized \acp{dnn}, which often require large data sets for training, though usually not as much as the generic \acp{dnn}. Furthermore, deep unfolding typically requires a high level of domain knowledge, such as explicit knowledge of the statistical model $\Pdf{\myVec{Y} | \myVec{S}}$ up to possibly some missing parameters, in order to formulate the optimization algorithm in a manner that can be unfolded. 

\subsection{Model-Based Deep Learning by \ac{dnn}-Aided Algorithms}
\label{subsec:ModelMLAlgorithms}
The second strategy for combining model-based methods and deep learning, which we refer to as {\em \ac{dnn}-aided hybrid algorithms}, aims at integrating \ac{ml} into model-based techniques. Such \ac{dnn}-aided systems mainly utilize conventional model-based methods for inference,  while incorporating \acp{dnn} to make the resultant system more robust and model-agnostic. This approach builds upon the insight that model-based algorithms typically consist of a set of generic manipulations that are determined by the {\em structure of the statistics}, e.g., whether it obeys a Markovian structure. Beside these generic manipulations, there are also computations that require actual knowledge of  $\Pdf{\myVec{Y} | \myVec{S}}$, as illustrated in Fig. \ref{fig:ModelBasedFlow}(a). Consequently, when one has prior knowledge on the structure of the underlying distribution but not of its actual distribution, \ac{ml} can be utilized to fill in the missing components required to carry out the algorithm.  

In particular, \ac{dnn}-aided hybrid algorithms start with a model-based algorithm that is suitable for inference when the statistics of $\Pdf{\myVec{Y} | \myVec{S}}$ are available. For instance, symbol detection over finite-memory channels can be carried out accurately and with affordable complexity using either the Viterbi algorithm \cite{viterbi1967error} or the BCJR method \cite{bahl1974optimal}, assuming $\Pdf{\myVec{Y} | \myVec{S}}$ is known. 
Then, \ac{ml}-based techniques, such as dedicated \acp{dnn}, are used to estimate only $\Pdf{\myVec{Y} | \myVec{S}}$ from data. These dedicated \acp{dnn} can be trained individually,  {separately from the inference task, or in an end-to-end manner along with the overall algorithm that maps $\myVec{Y}$ into an estimate of $\myVec{S}$}. An illustration of a \ac{dnn}-aided algorithm obtained by integrating \ac{ml} into the iterative methods illustrated in Fig. \ref{fig:ModelBasedFlow}(a) is depicted in Fig. \ref{fig:ModelBasedFlow}(c).

\ac{dnn}-aided hybrid algorithms have several  advantages: First, they use \acp{dnn} for specific intermediate tasks, such as computing a conditional probability measure, which are much simpler compared to end-to-end inference. Consequently, relatively simple networks that are trainable using small training sets can be used. Furthermore,  once trained the system effectively implements the model-based algorithm in a data-driven manner without imposing a model on the underlying distribution and estimating its parameters. Concrete examples of \ac{dnn}-aided symbol detection algorithms are detailed in Section \ref{subsec:SymbolDetectionAlgorithm}.

\section{Deep Symbol Detection}
\label{sec:SymbolDetection} 
In digital communication systems, the receiver is required to reliably recover the transmitted symbols from the observed channel output. This task is commonly referred to as {\em symbol detection}. 
In this section, we present how the strategies for combining \ac{ml} and model-based algorithms detailed in the previous section can be applied for data-driven symbol detection. We first formulate the symbol detection in Section \ref{subsec:SymbolDetectionProblem}, after which we discuss the applications of data-driven receivers based on conventional \ac{dnn} architectures, deep unfolding, and \ac{dnn}-aided algorithms,  in Sections \ref{subsec:SymbolDetectionEstablished}-\ref{subsec:SymbolDetectionAlgorithm}, respectively. For each strategy we begin with the main rationale behind this approach, present at least one concrete example, and discuss its pros and cons. Finally, we numerically compare the data-driven receivers to their model-based counterparts in Section \ref{subsec:SymbolDetectionSims}.

\subsection{The Symbol Detection Problem}
\label{subsec:SymbolDetectionProblem}
\begin{figure}
	\centering
	\includefig{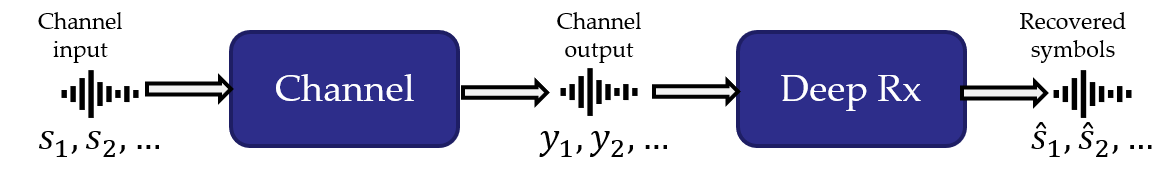} 
	\caption{Symbol detection illustration.} 
	\label{fig:SymDet2}
\end{figure}

To formulate the symbol detection problem, we let $\myS_i \in \mySet{S}^\Nusers$ be the symbol transmitted at time index $i \in \{1,2,\ldots, \Blklen\}:= \Blkset$.  Here, $\Blklen$ represents the blocklength and $\Nusers$ denotes the number of symbols transmitted at each time instance, e.g., the number of users transmitting simultaneously in the uplink communications channel.  Each symbol is uniformly distributed over a set of $\CnstSize$ constellation points, thus $| \mySet{S}| = \CnstSize$. We use $\myY_i\in \mySet{Y}^{\Nantennas}$ to denote the channel output at time index $i$, where $\Nantennas$ represents the number of receive antennas. When both $\Nantennas$ and $\Nusers$ are larger than one, the resulting setup corresponds to \ac{mimo} communications. Symbol detection refers to the recovery of $\myS^{\Blklen} := \{\myS_i\}_{i\in \Blkset}$ from the observed $\myY^{\Blklen}:= \{\myY_i\}_{i\in \Blkset}$. An illustration of the symbol detection problem using a \ac{dnn}-aided receiver is depicted in Fig.~\ref{fig:SymDet2}.

We focus on finite-memory stationary causal channels, where each  $\myY_i $ is given by a stochastic mapping of $\{\myS_l\}_{l=i-\Mem+1}^{i}$, and $\Mem$ is the memory of the channel, assumed to be smaller than the blocklength $\Blklen$. The special case in which $\Mem = 1$ is referred to as {\em flat} or  memoryless channel conditions.
The conditional \ac{pdf} of the channel output given its input thus satisfies   
\begin{equation}
\label{eqn:ChModel1}
\Pdf{\myY^{ \Blklen} | \myS^{ \Blklen}}\left(\myVec{y}^{ \Blklen} | \myVec{s}^{ \Blklen} \right)  = 
\prod\limits_{i\!=\!1}^{\Blklen}\Pdf{\myY_i | \{\myS_l\}_{l=i-\Mem+1}^{i}}\left(\myVec{y}_i  |\{\myVec{s}_l\}_{l=i-\Mem+1}^{i}\right),  
\end{equation}
where the lower-case $\myVec{y}_i$ and $\myVec{s}_i$ represent the realizations of the random variables $\myY_i$ and $\myS_i$, respectively. The fact that the channel is stationary implies that   the conditional \ac{pdf} $\Pdf{\myY_i | \{\myS_l\}_{l=i-\Mem+1}^{i}}$ does not depend on the  index $i$. 

The symbol detection mapping that minimizes the error rate is the \ac{map} rule, given by
\begin{align}
\hat{\myVec{s}}_i &= \mathop{\arg \max}\limits_{\myVec{s} \in \mySet{S}^\Nusers }\Pdf{\myS_i|\myVec{Y}^\Blklen}(\myVec{s}|\myVec{y}^\Blklen), \qquad i \in \Blkset. 
\label{eqn:MAP}
\end{align} 
For the memoryless case $\Mem=1$,  solving \eqref{eqn:MAP} reduces to maximizing 
$\Pdf{\myS|\myY}(\myVec{s}|\myVec{y}_i)$ over $\myVec{s} \in \mySet{S}^\Nusers$. However, when $\Nusers$ is large, as is commonly the case in uplink \ac{mimo} systems, solving \eqref{eqn:MAP} may be computationally infeasible, even when the \ac{pdf} $\Pdf{\myS|\myY}$ is perfectly known. The application of deep learning for symbol detection thus has two main motivations: the first is to allow symbol detection to operate in a model-agnostic manner, i.e., without requiring knowledge of  $\Pdf{\myS|\myY}$, and the second is to facilitate inference when the computational complexity of \eqref{eqn:MAP} renders solving it infeasible.

\subsection{Symbol Detection via Established Deep Networks}
\label{subsec:SymbolDetectionEstablished}
%

The first approach to designing data-driven symbol detectors treats the channel as a black-box and relies on well-known deep learning  architectures used in computer vision, speech, and language processing for detection.  We now describe how conventional deep neural architectures can be used for symbol detection.

\subsubsection{Overview of Design Process}
Different \ac{dnn} architectures have shown promising results for detection and estimation in applications such as image processing \cite{tian2018deep, wang2020deep, minaee2020image}, speech recognition \cite{hin12, gra14towards, amo16deep}, machine translation \cite{bah14,cho14learning,yang2020survey}, and bioinformatics \cite{li16, lan2018survey}.  Some of these neural network architectures can be used to design a symbol detector for channels with unknown models using supervised learning.  This process typically consists of the following steps:

\begin{enumerate}
	\item  Identify the conventional neural network architectures that are suitable for the channel under consideration, and use these networks as building blocks for designing the detection algorithm. For example, \acp{rnn} are more suitable for sequential detection in channels with memory, while convolutional and fully-connected networks are more suitable for memoryless channels.   
	\item Next, use channel input-output pairs to train the network. Two approaches can be used for training: In the first approach, a model is trained for each channel condition (e.g., each SNR). In the second approach, a large training dataset consisting of various channel conditions is used to train a single neural network detector for detection over a wide-range of channel conditions.  The training data can be generated by randomizing the transmitted symbols  and generating the corresponding received signal using mathematical models, simulations, experiments, or field measurements. 
	\item Train the overall resulting network in an end-to-end fashion.
\end{enumerate}

We next demonstrate how this rationale is translated into a concrete data-driven symbol detector architecture for finite-memory channels.

\subsubsection{Example: SBRNN for Finite-Memory Channels}

The \ac{sbrnn} is a sequence detection algorithm for finite-memory channels proposed in \cite{farsad2018neural}. Generally, sequence detection can be performed using \acp{rnn} \cite{lecun2015deep}, which are well established for sequence estimation in different problems such as  neural machine translation \cite{bah14}, speech recognition \cite{hinton2012deep}, or bioinformatics \cite{li16}.  For simplicity, we assume in our description that the input cardinality $\Nusers$ is $\Nusers = 1$.
The estimated symbol in this case is given by
\begin{align}
\label{eq:estForwSeqPMF}
\hat{\myVec{s}}_i &= \mathop{\arg \max}\limits_{\myVec{s}_i \in \mySet{S}}P_{\text{RNN}}(\myVec{s}_i|\myVec{y}^i), \qquad i \in \Blkset,
\end{align}    
where $P_{\text{RNN}}$ is the probability of estimating each symbol based on the \ac{dnn} model used. 
One of the main benefits of this detector is that after training, it can perform detection on any data stream as it arrives at the receiver. This is because the observations from previous symbols are summarized as the state of the RNN, which is represented by a vector. Note that the observed signal during the \jth~transmission, $\myVec{Y}_j$ where $j>k$, may carry information about the \kth~symbol $\myS_k$ due to the memory of the channel. However, since RNNs are feed-forward only, during the estimation of $\hat{\myS}_k$, the observation signal $\myVec{Y}_j$ is not considered.  

One way to overcome this limitation is by using bidirectional RNNs (BRNNs). In such networks, a sequence of $\BRNNLen$ received signals are  fed once in the forward direction into one RNN cell, and  fed  once in the backward direction into another RNN cell \cite{sch97}, for some fixed $\BRNNLen$ representing the BRNN length. The two outputs are then concatenated and may be passed to more bidirectional layers. 
A signal whose blocklength $\Blklen$ is larger than the BRNN length $\BRNNLen$ is divided into multiple distinct subsequences of length $\BRNNLen$. Ideally, the $\BRNNLen$ must be at least the same size as the memory length $\Mem$. However, if this is not known in advance, the BRNN length can be treated as a hyperparameter to be tuned during training. At time instance $i$ belonging to the \kth~subsequence,  the estimated symbol for BRNN is given by
\begin{align}
\label{eq:estBiDirSeqPMF}
\hat{\myVec{s}}_i &= \mathop{\arg \max}\limits_{\myVec{s}_i \in \mySet{S}}P_{\text{BRNN}}(\myVec{s}_i|\myVec{y}_{k}^{k+\BRNNLen-1}), \qquad k-\BRNNLen+1 \leq i \leq k.
\end{align}
To simplify the notation, we use $\hat{\myVec{p}}_i^{(k)}$ to denote the $\CnstSize\times 1$ matrix whose entries are the \ac{pmf} $P_{\text{BRNN}}(\myVec{s}_i|\myVec{y}_{k}^{k+\BRNNLen-1})$ for each $\myVec{s}_i \in \mySet{S}$. 

\begin{figure}
	\centering
	\includefig{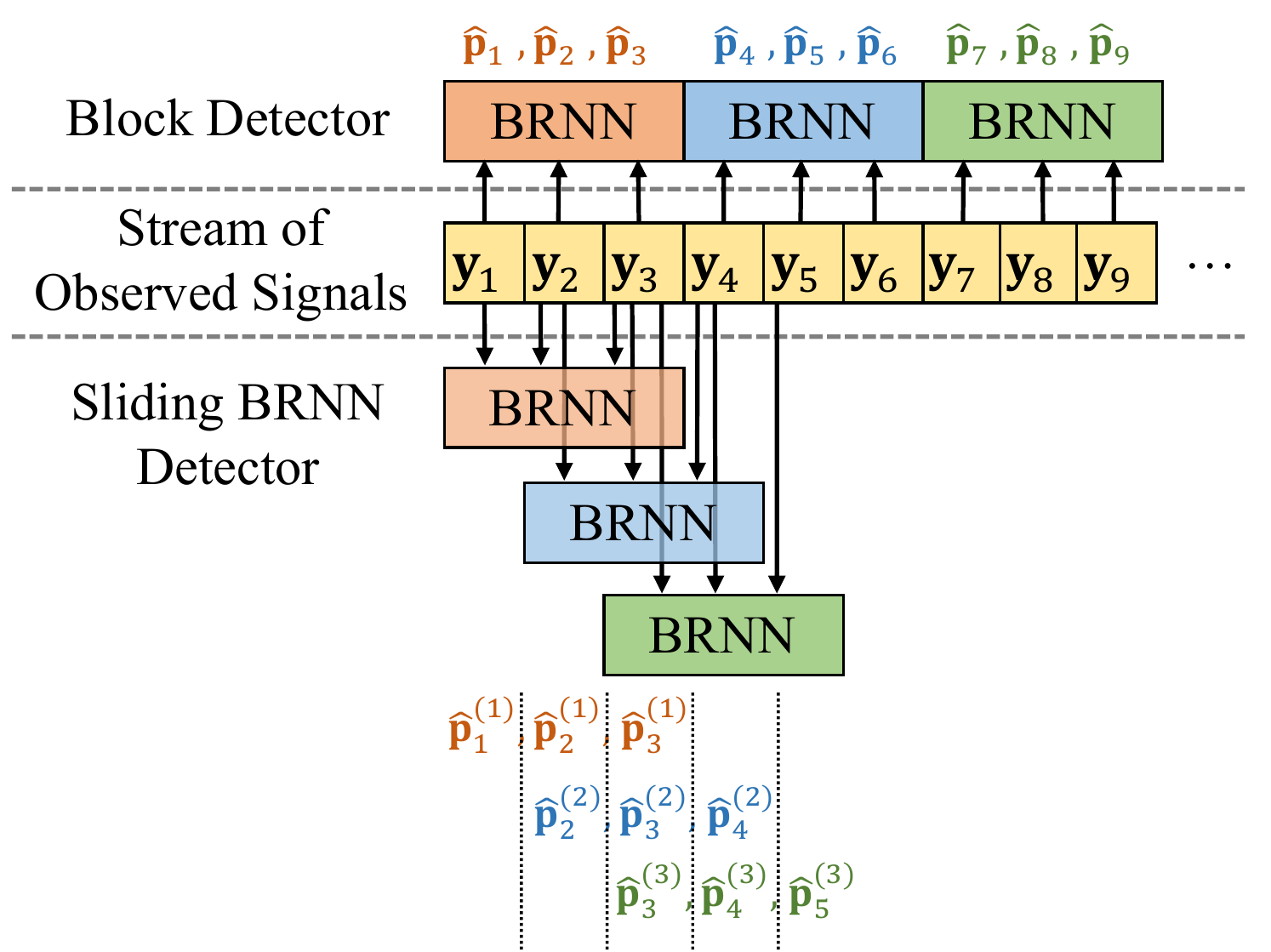}
	\caption{\label{fig:slidingDetector} The SBRNN detector versus using a BRNN as a block detector.}
	\vspace{-0.3cm}
\end{figure}

The BRNN architecture ensures that in the estimation of a symbol, future signal observations are taken into account. 
During training, blocks of $\BRNNLen$ consecutive transmissions are used for training.  
Once the network is trained, BRNNs detect the stream of incoming data in fixed blocks of length $\BRNNLen$, as shown in the top portion of Fig.~\ref{fig:slidingDetector}. The main drawback here is that the symbols at the end of each block may affect the symbols in the next block, and since each block is treated independently, this relation is not captured in this scheme. Another issue is that the block of  $\BRNNLen$ symbols must be received before detection can be performed. The top portion of Figure \ref{fig:slidingDetector} shows this scheme for $\BRNNLen=3$.

To overcome these limitations, in the \ac{sbrnn} proposed in \cite{farsad2018neural}, the first $\BRNNLen$ symbols are detected using the BRNN. Then, as each new symbol arrives at the receiver, the subsequence processed by the BRNN slides ahead by one symbol. Let the set $\mathcal{K}_i = \{k \mid k \leq i ~\wedge~ k + \BRNNLen > i \}$ be the set of all valid starting positions for a BRNN detector of length $\BRNNLen$, such that the detector overlaps with the \ith~symbol. For example, if $\BRNNLen=3$ and $i=4$, then $k=1$ is not in the set $\mathcal{K}_i$ since the BRNN detector overlaps with symbol positions 1, 2, and 3, and not the symbol position 4. 
The estimated \ac{pmf} corresponding to the \ith~symbol is given by the weighted sum of the estimated PMFs for each of the relevant windows:
\begin{align}
\label{eq:slidingPMF}
\hat{\myVec{p}}_i =  \sum_{k\in \mathcal{K}_i} \alpha_k \hat{\myVec{p}}_i^{(k)}, \qquad \alpha_k \geq 0 ~\text{and}~ \sum \alpha_k = 1.
\end{align}
The weighted sum coefficients can be set to  $\alpha_k = \frac{1}{|\mathcal{K}_i|}$, as we do in the numerical evaluations in Section \ref{subsec:SymbolDetectionSims}.
An illustration of the operation of the SBRNN detector based on \eqref{eq:slidingPMF} is depicted in the bottom portion of Fig.~\ref{fig:slidingDetector}. 



\subsubsection{Summary}
Well-known \ac{dnn} architectures can be trained in an end-to-end manner to perform symbol detection. This approach builds upon the success of existing model-agnostic \ac{dnn} structures, resulting in symbol detectors operating  without any knowledge about the underlying channel models. Furthermore, this strategy allows combining some basic level of domain knowledge in the selection of the architecture as well as the preparation of its input. For example, the SBRNN detector detailed above identifies the BRNN architecture as one that is capable of handling temporal correlation in finite-memory channels, while using a sliding subsequence to overcome some of the limitations of BRNN architectures when applied to different blocks independently. Also, \ac{dnn}s can well-capture the nonlinearities that may exist in the channel. For example, the \ac{sbrnn} was used as an autoencoder in \cite{karanov2019end, karanov2020experimental} to achieve state-of-the-art performance over optical channels, outperforming model-based nonlinear equalizers such as Voltera. Finally, these networks can also be computationally more efficient than the optimal maximum-likelihood sequence detector over finite-memory channels, specifically for channels with long memory.  
 
 The main drawback in using established deep networks for end-to-end symbol detection is that such architectures typically have a very large number of parameters, and thus require massive data sets for training. This renders online training using pilot sequences impractical.  {Moreover, when they are trained using data from a large set of channel conditions, the resulting network will not perform optimally for each of those channel conditions individually. Furthermore, even when the data set is extremely large and diverse, it is not likely to capture all expected channel conditions.} Finally, 
 conventional \ac{dnn} architectures are treated as black-boxes, and are in general not interpretable, making it difficult to come up with performance guarantees.  

\subsection{Symbol Detection via Deep Unfolding}
\label{subsec:SymbolDetectionUnfolding}
Unlike conventional \acp{dnn}, which utilize established architectures, in deep unfolding the network structure is designed following a model-based algorithm. We next describe how this model-based \ac{ml} technique can be applied for symbol detection, and detail a concrete example for flat Gaussian \ac{mimo} channels.

\subsubsection{Overview of Design Process}
Deep unfolding is a method for converting an iterative algorithm into a \ac{dnn} by designing each layer of the network to resemble a single iteration. As such, the rationale in applying deep unfolding  consists of the following steps:
\begin{enumerate}
	\item Identify an iterative optimization algorithm that is useful for the problem at hand. For instance, recovering the \ac{map} symbol detector for flat \ac{mimo} channels can be tackled using various iterative optimization algorithms, such as projected gradient descent, unfolded into DetNet  \cite{samuel2019learning}, as  described in the sequel.
	\item Fix a number of iterations in the optimization algorithm.
	\item Design the layers to initiate the operation of each iteration in a trainable fashion, as illustrated in Fig. \ref{fig:ModelBasedFlow}. 
	\item Train the overall resulting network in an end-to-end fashion.
\end{enumerate}
We next demonstrate how this rationale is translated into a concrete data-driven symbol detector architecture for flat \ac{mimo} channels, i.e., \eqref{eqn:ChModel1} with $\Mem=1$.

\subsubsection{Example: DetNet for Flat MIMO Channels}
DetNet is a deep learning based symbol detector proposed in \cite{samuel2019learning} for flat Gaussian \ac{mimo} channels. To formulate DetNet, we first detail the specific channel model for which it is designed, and then show how it is obtained by unfolding the projected gradient descent method for recovering the \ac{map} estimate.

\paragraph{Flat Gaussian MIMO Channel}
As we focus on stationary memoryless channels, we drop the subscript $i$ representing the time instance, and write the input-output relationship of a flat Gaussian \ac{mimo} channel as
\begin{equation}
\label{eqn:Gaussian}
\myY = \myMat{H}\myS + \myVec{W},
\end{equation}
where $\myMat{H}$ is a known deterministic $\Nantennas\times \Nusers$ channel matrix, and $\myVec{W}$ consists of $\Nantennas$ i.i.d Gaussian \acp{rv}. Consider the case in which the symbols are generated from a \ac{bpsk} constellation in a uniform i.i.d. manner, i.e., $\mathcal{S}=\{\pm 1\}$. In this case the \ac{map} rule in \eqref{eqn:MAP} given an observation $\myY=\myVec{y}$ becomes the minimum distance estimate, given by 
\begin{equation}
\label{eqn:DetNetObj}
\hat{\myVec{s}} = \mathop{\arg \min}\limits_{\myVec{s}\in\{\pm 1\}^\Nusers} \|\myVec{y}-\myMat{H}\myVec{s}\|^2.
\end{equation}

\paragraph{Project Gradient Descent Optimization}
While directly solving \eqref{eqn:DetNetObj} involves an exhaustive search over the $2^\Nusers$ possible symbol combinations, it can be tackled with affordable computational complexity using the iterative projected gradient descent algorithm. Let $\mySet{P}_{\mySet{S}}(\cdot)$ denote the projection into the $\mySet{S}$ operator, which for \ac{bpsk} constellations is the sign function. The projected gradient descent iteratively refines its estimate, which at iteration index $q+1$ is obtained recursively as
\begin{align}
\hat{\myVec{s}}_{q + 1} 
&= \mySet{P}_{\mySet{S}}\left(\hat{\myVec{s}}_{q} - \eta_q \left.\frac{\partial \|\myVec{y}-\myMat{H}\myVec{s}\|^2}{\partial \myVec{s}}\right|_{\myVec{s} = \hat{\myVec{s}}_q} \right) \notag \\
&=  \mySet{P}_{\mySet{S}}\left(\hat{\myVec{s}}_{q} - \eta_q\myMat{H}^T\myVec{y} + \eta_q\myMat{H}^T\myMat{H}\hat{\myVec{s}}_{q} \right),
\label{eqn:ProjGrad}
\end{align}
where $\eta_q$ denotes the step size at iteration $q$, and $\hat{\myVec{s}}_0$ is set to some initial guess.  

\paragraph{Unfolded DetNet}
DetNet unfolds the projected gradient descent iterations in \eqref{eqn:ProjGrad} into a \ac{dnn}, which learns to carry out this optimization procedure from data. To formulate DetNet, we first fix a number of iterations $\Niter$. Next, we design a  \ac{dnn} with $\Niter$ layers, where each layer imitates a single iteration of \eqref{eqn:ProjGrad} in a trainable manner.

\begin{figure}
	\centering
	\includegraphics[width=\linewidth]{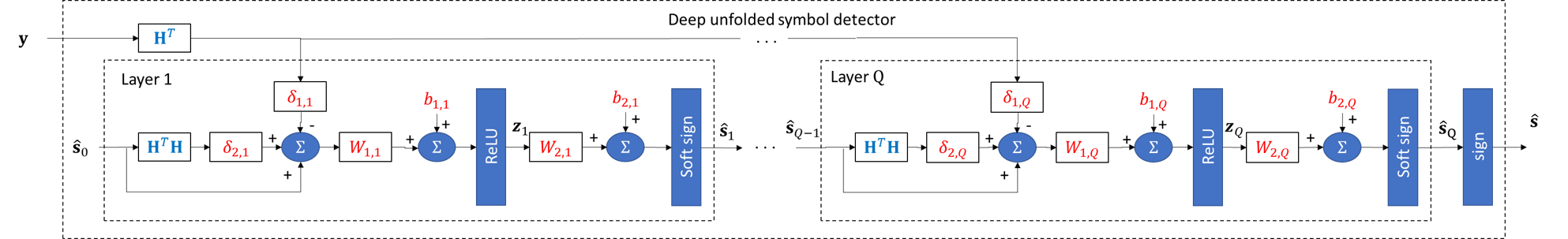} 
	\caption{Deep unfolded symbol detector illustration. Parameters in red fonts are learned in training, while those in blue fonts are externally provided.} 
	\label{fig:DetNet}
\end{figure}

In particular, DetNet builds upon the observation that each projected gradient descent iteration consists of two stages: gradient descent computation, i.e., $\hat{\myVec{s}}_{q} - \eta_q\myMat{H}^T\myVec{y} + \eta_q\myMat{H}^T\myMat{H}\hat{\myVec{s}}_{q}$, and projection, namely, applying $ \mySet{P}_{\mySet{S}}(\cdot)$. Therefore, each unfolded iteration is represented as two sub-layers: The first sub-layer learns to compute the gradient descent stage by treating the step-size as a learned parameter and applying a conventional fully-connected layer with ReLU activation to the obtained value. For iteration index $q$, this results in 
\begin{equation}
\myVec{z}_{q} = {\rm ReLU}\left(\myMat{W}_{1,q}\left(\hat{\myVec{s}}_{q-1} - \delta_{1,q}\myMat{H}^T\myVec{y} + \delta_{2,q}\myMat{H}^T\myMat{H}\hat{\myVec{s}}_{q-1}  \right) + \myVec{b}_{1,q}  \right),
\label{eqn:Layer1}
\end{equation}
in which $\{\myMat{W}_{1,q}, \myVec{b}_{1,q}, \delta_{1,q}, \delta_{2,q}\}$ are learnable parameters. 
The second sub-layer learns the projection operator by approximating the sign operation with a soft sign activation proceeded by a fully-connected layer, leading to
\begin{equation}
\hat{\myVec{s}}_q = {\rm soft~sign}\left( \myMat{W}_{2,q} \myVec{z}_{q} +\myVec{b}_{2,q} \right).
\label{eqn:Layer2}
\end{equation}
Here, the learnable parameters are $\{\myMat{W}_{2,q}, \myVec{b}_{2,q}\}$. 
The resulting deep network is depicted in Fig. \ref{fig:DetNet}, in which $\hat{\myVec{s}}_0 $ is set to some initial guess, and the output after $\Niter$ iterations, denoted $\hat{\myVec{s}}_{\Niter}$, is used as the estimated symbol vector by taking the sign of each element.

Let  $\myVec{\theta} = \{\myMat{W}_{1,q}, \myMat{W}_{2,q}, \myVec{b}_{1,q}, \myVec{b}_{2,q}, \delta_{1,q}, \delta_{2,q}\}_{q=1}^Q$ be the trainable parameters of DetNet$^{\footnotemark}$.
\footnotetext{The formulation of DetNet in \cite{samuel2019learning} includes an additional sub-layer in each iteration intended to further lift its input into higher dimensions and introduce additional trainable parameters, as well as reweighing of the outputs of subsequent layers. As these operations do not follow directly from unfolding the projected gradient descent method, they are not included in the description here.} To tune $\myVec{\theta}$, the overall network is trained in an end-to-end manner to minimize the empirical weighted $\ell_2$ norm loss over its intermediate layers. In particular, by letting $\{\myVec{s}_t, \myVec{y}_t\}_{t=1}^{\Ntraining}$ denote the training set consisting of channel outputs and their corresponding transmitted symbols, the loss function used for training  DetNet is given by
\begin{equation}
\label{eqn:LossDetNet}
\mySet{L}(\myVec{\theta}) = \frac{1}{\Ntraining}\sum_{t=1}^{\Ntraining}\sum_{q=1}^{\Niter} \log(q)\|\myVec{s}_t - \hat{\myVec{s}}_q(\myVec{y}_t; \myVec{\theta}) \|^2,
\end{equation}
where $\hat{\myVec{s}}_q(\myVec{y}_t; \myVec{\theta})$ is the output of the $q$th layer of DetNet with parameters $\myVec{\theta}$ and input $\myVec{y}_t$. This loss measure accounts for the interpretable nature of the unfolded network, in which the output of each layer is a further refined estimate of $\myVec{S}$.

\subsubsection{Summary}
Deep unfolding incorporates model-based domain knowledge to obtain a dedicated \ac{dnn} design, which follows an iterative optimization algorithm. Compared to the conventional \acp{dnn} discussed in the previous section, unfolded networks are typically interpretable, and tend to have a smaller number of parameters, and can thus be trained quicker \cite{balatsoukas2019deep}. Nonetheless, these deep networks are still  highly parameterized, and require a large volume of training data. For instance, DetNet is trained in \cite{samuel2019learning} using approximately $250$ million labeled samples. 

One of the key properties of unfolded networks is their reliance on model knowledge. For example, the unfolded receiver must know that the channel input-output relationship takes the form \eqref{eqn:Gaussian} in order to formulate the projected gradient iterations \eqref{eqn:ProjGrad}, which in turn are unfolded into DetNet. 
The model-awareness of deep unfolding has its advantages and drawbacks. When the model is accurately known, deep unfolding essentially incorporates it into the \ac{dnn} architecture, as opposed to conventional \acp{dnn} that must learn this from data. However, this approach does not exploit the model-agnostic nature of deep learning, and thus may achieve degraded performance when the true channel conditions deviate from the model assumed in design, e.g., \eqref{eqn:Gaussian}. 
In particular, a key advantage of deep unfolding over the model-based optimization algorithm is in inference speed. For instance, DetNet requires fewer layers to reliably detect compared to the number of iterations required for projected gradient descent to converge. 

{
Another important advantage of unfolded networks is their ability to improve the accuracy compared to the iterative optimization algorithm from which they originate. In particular, the set $\{\pm 1\}^{\Nusers}$ over which the optimization problem \eqref{eqn:DetNetObj} is formulated is not convex, and thus projected gradient descent is not guaranteed to recover its solution, regardless of the number of iterations. By unfolding the application of projected gradient descent for solving \eqref{eqn:DetNetObj} into a \ac{dnn} with trainable parameters, the resulting network is often able to overcome this difficulty and converge to the true solution of the optimization problem when properly trained, despite the non-convexity.
Finally, in the context of receiver design, most unfolded networks to date, including DetNet as well as OAMP-net \cite{he2018model} which unfolds the orthogonal approximate message passing optimization algorithm, require \ac{csi}. For example, knowledge of the matrix $\myMat{H}$ is utilized in the architecture depicted in Fig.~\ref{fig:DetNet}. This implies that additional mechanisms for estimating the channel must be incorporated into the receiver architecture \cite{he2019model}. Nonetheless, while the aforementioned deep unfolding based receivers require \ac{csi}, unfolded networks can be designed without such knowledge \cite{monga2019algorithm}. For example, one can unfold the optimization algorithm assuming \ac{csi} is available, and then treat $\myMat{H}$ which appears in the unfolded network as part of its trainable parameters.}

\subsection{Symbol Detection via \ac{dnn}-Aided Algorithms}
\label{subsec:SymbolDetectionAlgorithm}
\ac{dnn}-aided hybrid algorithms combine domain knowledge in the form of a model-based inference algorithm for the problem at hand. This strategy allows the design of model-based \ac{ml} systems with varying levels of domain knowledge, in which deep learning is used to robustify and remove model-dependence of specific components of the algorithm. In the following we first review the rationale when designing \ac{dnn}-aided symbol detectors, after which we detail three concrete examples arising from different symbol detection algorithms. 
 
\subsubsection{Overview of Design Process}
\ac{dnn}-aided algorithms aim to carry out model-based inference methods in a data-driven fashion. These hybrid systems thus utilize deep learning not for the overall inference task, but for robustifying and relaxing the model-dependence of established model-based inference algorithms. Consequently, the design of \ac{dnn}-aided hybrid systems consists of the following steps:
\begin{enumerate}
	\item First, a proper inference algorithm is chosen. In particular, the domain knowledge is encapsulated in the selection of the algorithm that is learned from data. For example, the Viterbi algorithm is a natural candidate for symbol detection over finite-memory channels when seeking a symbol detector capable of operating in real-time, or alternatively, the BCJR scheme is the suitable choice for carrying out \ac{map} inference over such channels. When designing receivers for flat \ac{mimo} channels, interference cancellation methods may be the preferable algorithmic approach for symbol detection. We show how these  methods are converted into \ac{dnn}-aided algorithms in the sequel. 
	\item Once a model-based algorithm is selected, we identify its model-specific computations, and replace them with dedicated compact \acp{dnn}.
	\item The resulting \acp{dnn} can be either trained individually, or the overall system can be trained in an end-to-end manner.
\end{enumerate}

Since the implementation of \ac{dnn}-aided algorithms highly varies with the selection of the learned model-based method, we next present three concrete examples in the context of symbol detection: {\em ViterbiNet},  which learns to carry out Viterbi detection \cite{viterbi1967error}; {\em BCJRNet}, which implements the BCJR algorithm of \cite{bahl1974optimal} in a data-driven fashion; and {\em DeepSIC}, which is based on the soft iterative interference cancellation methods for \ac{mimo} symbol detection \cite{choi2000iterative}.

\subsubsection{Example: ViterbiNet for Finite-Memory Channels}
ViterbiNet proposed in \cite{shlezinger2019viterbinet} is a data-driven implementation of the Viterbi detection algorithm \cite{viterbi1967error}, which is one of the most common workhorses in digital communications.  This \ac{dnn}-aided symbol detection algorithm is suitable for finite-memory channels of the form \eqref{eqn:ChModel1}, without requiring prior knowledge of the channel conditional distributions $\Pdf{\myY_i | \{\myS_l\}_{l=i-\Mem+1}^{i}}$. For simplicity, we assume in our description that the input cardinality $\Nusers$ is $\Nusers = 1$. As a preliminary step to presenting ViterbiNet, we now briefly review conventional model-based Viterbi detection.

\paragraph{The Viterbi Algorithm}
The Viterbi algorithm recovers the maximum likelihood sequence detector, i.e.,
\begin{align}
\hat{\myVec{s}}^{ \Blklen}\left( \myVec{y}^{ \Blklen}\right)  
&:= \mathop{\arg \max}_{\myVec{s}^{ \Blklen} \in \mySet{S}^\Blklen } \Pdf{\myVec{Y}^{ \Blklen} | \myVec{S}^{ \Blklen}}\left( {\myVec{y}^{ \Blklen} | \myVec{s}^{ \Blklen}}\right)  \notag \\
&
= \mathop{\arg \min}_{\myVec{s}^{ \Blklen} \in \mySet{S}^\Blklen } -\log \Pdf{\myVec{Y}^{ \Blklen} | \myVec{S}^{ \Blklen}}\left( \myVec{y}^{ \Blklen} | \myVec{s}^{ \Blklen}\right) .
\label{eqn:ML1}
\end{align}
Using \eqref{eqn:ChModel1} the optimization problem \eqref{eqn:ML1} becomes 
\begin{align}
\hat{\myVec{s}}^{ \Blklen}\left( \myVec{y}^{ \Blklen}\right)   
&= \mathop{\arg \min}_{\myVec{s}^{ \Blklen} \in \mySet{S}^\Blklen }\sum\limits_{i=1}^{\Blklen } - \log \Pdf{\myVec{Y}_i | \{\myVec{S}_l\}_{l=i-\Mem+1}^{i}}\left( \myVec{y}_i  | \{\myVec{s}_l\}_{l=i-\Mem+1}^{i}\right).
\label{eqn:ML3} 
\end{align}

To proceed, we define a state variable $\myState_i := [\myS_{i-\Mem+1},\ldots, \myS_i] \in \mySet{S}^\Mem$. Since the symbols are i.i.d. and uniformly distributed, it follows that $\Pdf{\myState_i|\myState_{i-1}}(\myStateR_i|\myStateR_{i-1}) = \CnstSize^{-1}$ when $\myStateR_i$ is a shifted version of $\myStateR_{i-1}$, i.e., the first $\Mem-1$ entries of $\myStateR_i$  are the last $\Mem-1$ entries of $\myStateR_{i-1}$, and zero otherwise. We can now write \eqref{eqn:ML3} as 
\begin{align}
\hat{\myVec{s}}^{ \Blklen}\left( \myVec{y}^{ \Blklen}\right)   
&= \mathop{\arg \min}_{\myVec{s}^{ \Blklen} \in \mySet{S}^\Blklen }\sum\limits_{i=1}^{\Blklen } - \log \Pdf{\myVec{Y}_i |\myState_i}\left( \myVec{y}_i  | \{\myVec{s}_l\}_{l=i-\Mem+1}^{i}\right).
\label{eqn:ML4} 
\end{align}

\begin{figure}
	\centering
	\includegraphics[width=\linewidth]{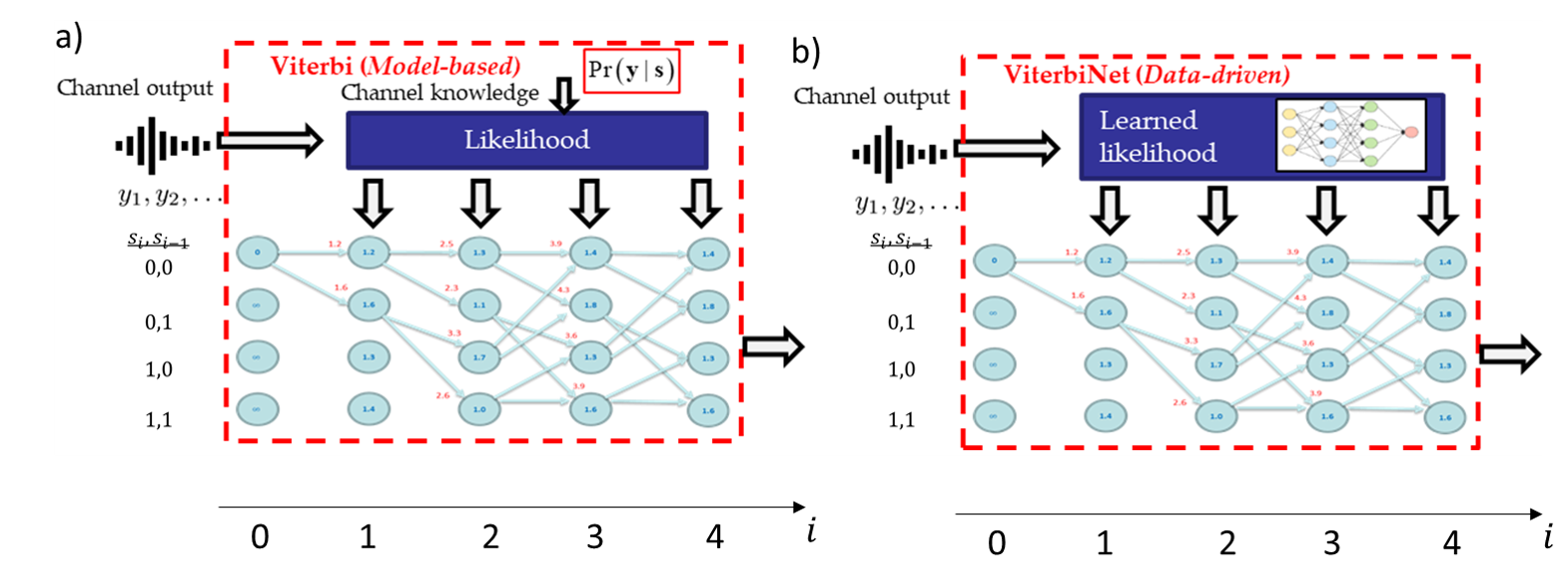} 
	\caption{An illustration of the operation of the Viterbi algorithm (a) and ViterbiNet (b). Here, the memory length is $\Mem=2$ and the constellation is $\mySet{S} = \{0,1\}$, implying that the number of states is $\CnstSize^\Mem = 4$. The values in blue are the negative log-likelihoods for each state, while the quantities in red represent the updated path loss $c_i(\cdot)$. When multiple paths lead to the same state, as occurs for $i= 3,4$, the path with minimal loss is maintained.} 
	\label{fig:ViterbiNet}
\end{figure}

The optimization problem \eqref{eqn:ML4} can be solved recursively using dynamic programming, by  iteratively updating a {\em path cost} $c_i(\myStateR)$ for each state $\myStateR\in \mySet{S}^{\Mem}$. The resulting scheme, known as the Viterbi algorithm, is given below as Algorithm~\ref{alg:Algo1}, and illustrated in Fig. \ref{fig:ViterbiNet}(a). 
\begin{algorithm}  
	\caption{ The Viterbi Algorithm \cite{viterbi1967error}}
	\label{alg:Algo1}
	\KwData{Fix an initial path $\myVec{p}_{0}\left(s\right) = \varnothing $ and path cost ${c}_{0}\!\left(\myStateR\right)\! =\!  0$,  $\myStateR \in \mySet{S}^{\Mem}$. }
	\For{$i=1,2,\ldots,\Blklen$}{
		For each state $\myStateR \in \mySet{S}^{\Mem}$, compute previous state with shortest path, denoted $u_{\myStateR}$, via 
		\begin{equation}
		\label{eqn:ViterbiUpdate}
		u_{\myStateR} = \mathop{\arg\min}\limits_{\myStateR' \in \mySet{S}^\Mem:  \Pdf{\myState_i|\myState_{i-1}}(\myStateR|\myStateR')>0 } \left({c}_{i-1}\left(\myStateR'\right)  - \log \Pdf{\myY_i  | \myState_i}\left( \myVec{y}_i  |\myStateR\right) \right).
		\end{equation}\\
		Update cost and path via 			
		\begin{equation}
		\label{eqn:ViterbiUpdate2}
		{c}_i\left( \myStateR\right) = {c}_{i-1}\left(u_{\myStateR} \right)  - \log \Pdf{\myY_i  | \myState_i}\left( \myVec{y}_i  | \myStateR\right), 
		\end{equation}
		and $\myVec{p}_{i}\left(\myStateR\right)  = \big[\myVec{p}_{i-1}\left(u_{\myStateR}\right) , u_{\myStateR} \big]$\;
	}
	\KwOut{$\hat{\myVec{s}}^\Blklen = \myVec{p}_{\Blklen}\left(\myStateR^*\right)$ where $\myStateR^* = \arg\min_{\myStateR} c_\Blklen(\myStateR)$.}
\end{algorithm}

The Viterbi algorithm has two major advantages: $1)$ It solves \eqref{eqn:ML1} at a computational complexity that is linear in the blocklength $\Blklen$. For comparison, the computational complexity of solving \eqref{eqn:ML1} directly grows exponentially with $\Blklen$;   	
$2)$ The algorithm produces estimates sequentially during run-time. In particular, while in \eqref{eqn:ML1}  the estimated output $\hat{\myVec{s}}^{ \Blklen}$ is computed using the entire received block $\myVec{y}^{ \Blklen}$, Algorithm~\ref{alg:Algo1} computes $\hat{\myVec{s}}_i$  once $\myVec{y}_{i + \Mem - 1}$ is received.  

\paragraph{ViterbiNet}
ViterbiNet proposed in \cite{shlezinger2019viterbinet} learns to implement  the Viterbi algorithm from data in a model-agnostic manner. Following the rationale of \ac{dnn}-aided algorithms, this is achieved by identifying the model-based components of the algorithm, which for  Algorithm~\ref{alg:Algo1} boils down to the computation of the log-likelihood function $ \log \Pdf{\myY_i  | \myState_i}\left( \myVec{y}_i  |\myStateR\right) $. 
Once  this quantity is computed for each $\myStateR \in \mySet{S}^\Mem$, the Viterbi algorithm only requires knowledge of the memory length $\Mem$. This requirement is much easier to satisfy compared to full \ac{csi}.  

Since the channel is stationary, it holds that the log-likelihood function   depends only on the realizations of $\myVec{y}_i$ and of $\myStateR$, and not on the time index $i$. 
Therefore, to implement Algorithm \ref{alg:Algo1} in a data-driven fashion, ViterbiNet replaces the explicit computation of the log-likelihoods with an \ac{ml}-based system that learns to evaluate this function from training data. In this case, the input of the system is the channel output realization $\myVec{y}_i$ and the output is an estimate of $ \log \Pdf{\myY_i  | \myState_i}\left( \myVec{y}_i  |\myStateR\right)$ for each $\myStateR\in\mySet{S}^{\Mem}$. The rest of the Viterbi algorithm remains intact, and the detector implements Algorithm~\ref{alg:Algo1} using the learned log-likelihoods.   The proposed architecture is illustrated in Fig.~\ref{fig:ViterbiNet}(b).

\begin{figure}
	\centering
	{\includegraphics[width=\linewidth]{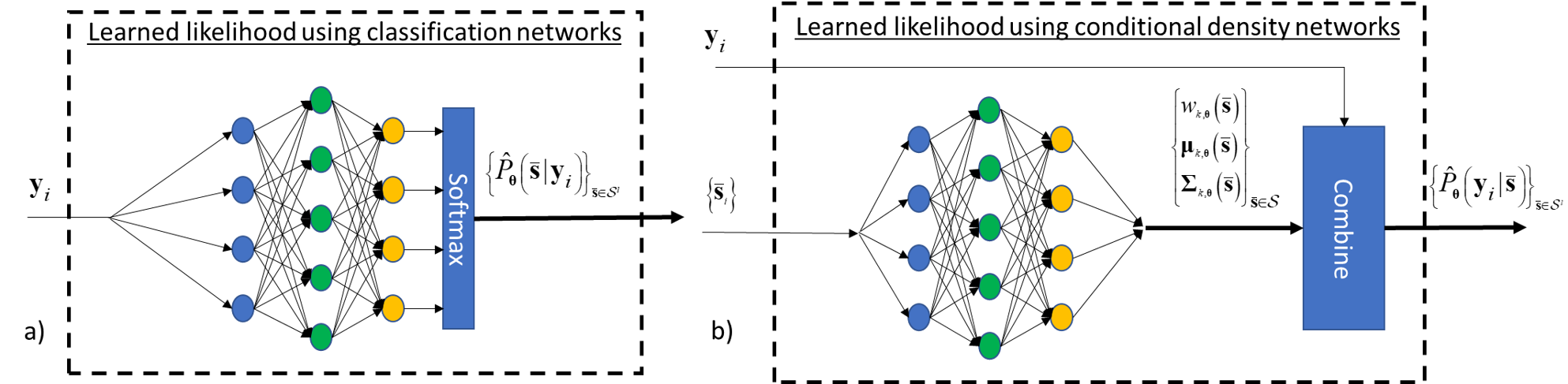}} 
	\caption{Learned likelihood architectures based on (a)  classification and  (b) conditional density estimation networks.} 
	\label{fig:LearnedFunctionNode}	 
\end{figure}

Two candidate  architectures are considered for learning to compute the log likelihood,  one based on classification networks and one using density estimation networks.
\\
{\bf Learned likelihood using classification networks:}   Since $\myVec{y}_i$ is given and may take continuous values while the desired variables  take discrete values, a natural approach to evaluate $\Pdf{\myY_i  | \myState_i}\left( \myVec{y}_i  |\myStateR\right)$ for  each $\myStateR\in\mySet{S}^{\Mem}$   is to estimate $\Pdf{\myState_i| \myY_i  }\left( \myStateR | \myVec{y}_i\right)$ and then use Bayes rule to obtain
\begin{equation}
\label{eqn:Bayes}
\Pdf{\myY_i  | \myState_i}\left( \myVec{y}_i  |\myStateR\right)  =
\Pdf{\myState_i| \myY_i  }\left( \myStateR | \myVec{y}_i\right) \Pdf{\myY_i  }\left( \myVec{y}_i    \right) \CnstSize^{\Mem}.
\end{equation}
A parametric estimate of $\Pdf{\myState_i| \myY_i  }\left( \myStateR | \myVec{y}_i\right)$, denoted $\hat{P}_{\myVec{\theta}}(\myStateR|\myVec{y}_i)$,  is obtained  for each $\myStateR\in\mySet{S}^{\Mem}$ by training classification networks with softmax output layers to minimize the cross entropy loss. 
Here, for a labeled set $\{\myVec{s}_t, \myVec{y}_t\}_{t=1}^{\Ntraining}$,  the loss function is
\begin{equation}
\label{eqn:LossClass}
\mySet{L}(\myVec{\theta}) = \frac{1}{\Ntraining}\sum_{t=1}^{\Ntraining} -\log \hat{P}_{\myVec{\theta}}([\myVec{s}_{t-\Mem+1},\ldots, \myVec{s}_t]|\myVec{y}_t).
\end{equation}

In general, the  marginal \ac{pdf} of $\myY_i$ can be estimated from the training data using mixture density estimation via, e.g., \ac{em}  \cite[Ch. 2]{mclachlan2004finite}, or any other finite mixture model fitting method. However, obtaining an accurate density estimation becomes challenging  when $\myY_i$ is high-dimensional. 
Since $ \Pdf{\myY_i  }\left(\myVec{y}_i  \right)$  does not depend on the variable $\myStateR$, setting $ \Pdf{\myY_i  }\left(\myVec{y}_i  \right) \equiv 1$  does not affect the decisions in Algorithm \ref{alg:Algo1} due to the $\arg\min$ arguments, which are invariant to scaling the conditional distribution by a term that does not depend on $\myStateR$. 
The resulting structure  is illustrated in  Fig.~ \ref{fig:LearnedFunctionNode}(a).
\\
{\bf Learned likelihood using conditional density networks:} An additional strategy is to directly estimate the conditional  $\Pdf{\myY_i  | \myState_i}\left( \myVec{y}_i  |\myStateR\right)$  from data. This can be achieved using conditional density estimation networks \cite{bishop1994mixture, rothfuss2019conditional}  that are specifically designed to learn such \acp{pdf}, or alternatively, using normalizing flow networks to learn complex densities \cite{kobyzev2019normalizing}. 
For example, mixture density networks \cite{bishop1994mixture} model the conditional \ac{pdf} $\Pdf{\myY_i  | \myState_i}\left( \myVec{y}_i  |\myStateR\right)$ as a Gaussian mixture, and train a \ac{dnn} to  estimate  its mixing parameters, mean values, and covariances, denoted $w_{k,\myVec{\theta}}(\myStateR )$,  $\myVec{\mu}_{k,\myVec{\theta}}( \myStateR)$ and $\myMat{\Sigma}_{k,\myVec{\theta}}(\myStateR)$, respectively, by maximizing the likelihood  $\hat{P}_{\myVec{\theta}}\left(\myVec{y}_i  |\myStateR  \right)= \sum_{k} w_{k,\myVec{\theta}}(\myStateR ) \mathcal{N}\big(\myVec{y}_i | \myVec{\mu}_{k,\myVec{\theta}}(\myStateR), \myMat{\Sigma}_{k,\myVec{\theta}}(\myStateR)\big)$,  
as illustrated in  Fig. \ref{fig:LearnedFunctionNode}(b).  

Both approaches can be utilized for learning to compute the likelihood in ViterbiNet. When the channel outputs are high-dimensional, i.e., $\Nantennas$ is large, directly learning the conditional density is difficult and likely to be inaccurate. In such cases, the classification-based architecture, which avoids the need to explicitly learn the density, 
may be preferable. When the state cardinality $\CnstSize^\Mem$ is large, conditional density networks are expected to be more reliable. However, the Viterbi algorithm becomes computationally infeasible when  $\CnstSize^\Mem$  grows, regardless of whether it is implemented in a model-based or data-driven fashion, making ViterbiNet non-suitable for such setups.

\subsubsection{Example: BCJRNet for Finite-Memory Channels}
Factor graph methods, such as the sum-product algorithm, exploit the factorization of a joint distribution to efficiently compute a desired quantity \cite{kschischang2001factor}. In particular, the application of the sum-product algorithm for the joint input-output distribution of finite-memory channels allows for computing the \ac{map} rule, an operation whose burden typically grows exponentially with the block size, with complexity that only grows linearly with $\Blklen$. This instance of the sum-product algorithm is exactly the BCJR  detector proposed \cite{bahl1974optimal}. In the following we show how the BCJR method can be extended into the \ac{dnn}-aided BCJRNet. As in our description of ViterbiNet, we again focus on the case of $\Nusers=1$, and begin by presenting the model-based BCJR algorithm.

\paragraph{The BCJR Algorithm}
The BCJR algorithm computes the \ac{map} rule in \eqref{eqn:MAP} for finite-memory channels with complexity that grows linearly with the block size. To formulate this method, we recall the definition of $\myState_i$, and define the function
\begin{align} 
f\left(\myVec{y}_i, \myStateR_i, \myStateR_{i-1} \right) &:=
\Pdf{\myY_i | \myState_i }\left(\myVec{y}_i|\myStateR_{i} \right) \Pdf{\myState_i |  \myState_{i-1}}\left(\myStateR_{i}| \myStateR_{i-1}\right) \notag \\
&=
\begin{cases}
\frac{1}{\CnstSize}\Pdf{\myY_i | \myState_i }\left(\myVec{y}_i|\myStateR_{i} \right)&  (\myStateR_i)_j = (\myStateR_{i\!-1})_{j\!-\!1}, \quad \forall j\in\{2,\ldots, \Mem\}, \\
0 & {\rm otherwise}. 
\end{cases}
\label{eqn:FSC_funcNode}
\end{align} 
Combining \eqref{eqn:FSC_funcNode} and \eqref{eqn:ChModel1}, we obtain a factorizable expression of the joint distribution $\Pdf{\myVec{Y}^\Blklen,\myVec{S}^\Blklen}(\cdot)$, given by
\begin{align}
\Pdf{\myY^{ \Blklen} , \myS^{ \Blklen}}\left(\myVec{y}^{ \Blklen} , \myVec{s}^{ \Blklen} \right)  
&= 
\prod\limits_{i\!=\!1}^{\Blklen}\frac{1}{\CnstSize}\Pdf{\myY_i | \myState_i}\left(\myVec{y}_i  |[\myVec{s}_{i-\Mem+1}, \ldots, \myVec{s}_{i}]\right) \notag \\
&=   \prod\limits_{i\!=\!1}^{\Blklen} f\left(\myVec{y}_i, [\myVec{s}_{i-\Mem+1}, \ldots, \myVec{s}_{i}], [\myVec{s}_{i-\Mem}, \ldots, \myVec{s}_{i-1}] \right).
\label{eqn:ChModel2}
\end{align}
The factorizable expression of the joint distribution \eqref{eqn:ChModel2} implies that it can be represented as a factor graph with $\Blklen$ function nodes $\{	f\left(\myVec{y}_i, \myStateR_i, \myStateR_{i-1} \right) \}$, in which $\{\myStateR_i\}_{i=2}^{\Blklen-1}$ are edges while the remaining variables are half-edges$^{\footnotemark}$.
\footnotetext{Here we use Forney style factor graphs \cite{forney2001codes}, where variables are represented as edges or half-edges. However, it is also possible to represent variables as variables notes.} 

%

%
Using its factor graph representation, one can compute the joint distribution of $\myVec{S}^{\Blklen}$ and $\myVec{Y}^{\Blklen}$  by recursive message passing along this factor graph. In particular, 
\begin{align}
\Pdf{\myState_k,\myState_{k+1}, \myVec{Y}^{\Blklen}}(\myStateR_k,\myStateR_{k+1}, \myVec{y}^{\Blklen}) &= \FwdMsg{f_k}{\myState_k}(\myStateR_k) f({y}_{k+1},   \myStateR_{k+1}, \myStateR_k) 
\BwdMsg{f_{k+2}}{\myState_{k+1}}(\myStateR_{k+1}),
\label{eqn:Recursion1}
\end{align}
where the forward path messages satisfy 
\begin{equation}
\FwdMsg{f_i}{\myState_i}(\myStateR_i) = \sum_{\myStateR_{i-1}} f(\myVec{y}_{i},  \myStateR_{i}, \myStateR_{i-1})\FwdMsg{f_{i-1}}{\myState_{i-1}}(\myStateR_{i-1}),
\label{eqn:Recursion1Forwards}
\end{equation}
for $i = 1, 2,\ldots, k$. Similarly, the backward messages are 
\begin{equation}
\BwdMsg{f_{i\!+\!1}}{ \myState_i}(\myStateR_i) = \sum_{\myStateR_{i\!+\!1}} f(\myVec{y}_{i\!+\!1}, \myStateR_{i\!+\!1}, \myStateR_{i})\BwdMsg{f_{i\!+\!2} }{ \myState_{i\!+\!1}}(\myStateR_{i\!+\!1}),
\label{eqn:Recursion1Backwards}
\end{equation}
\begin{figure}
	\centering
	{\includefig{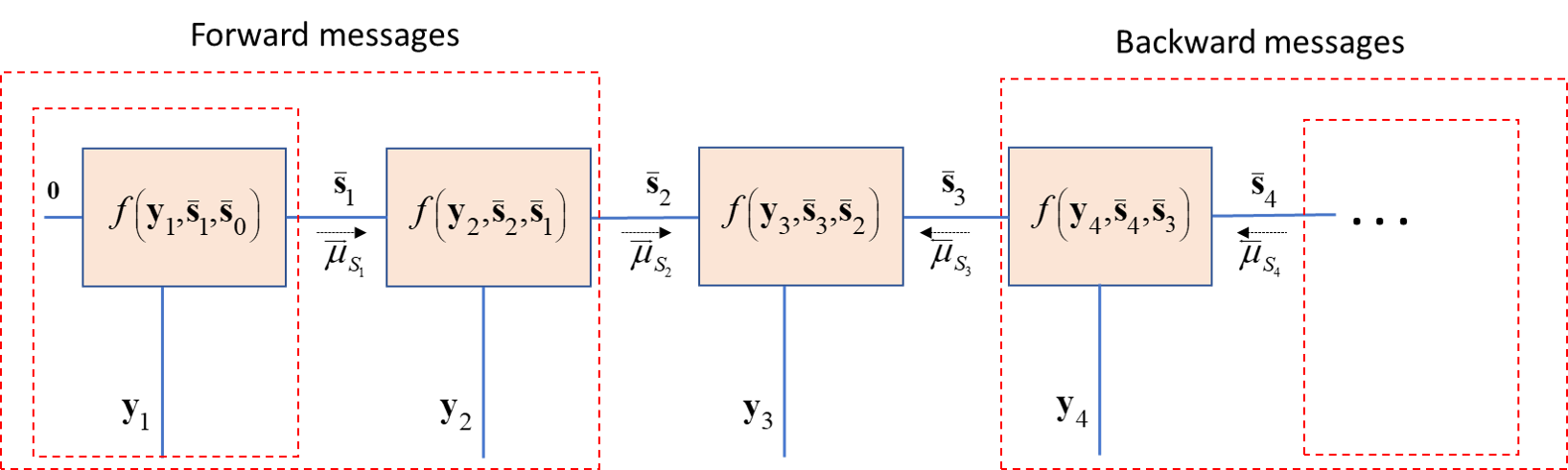}} 
	\caption{Message passing over the factor graph of a finite-memory channel.}
	\label{fig:SumProduct2}	 
\end{figure} 
for $i = \Blklen-1, \Blklen -2, \ldots, k+1$. This  message passing is  illustrated in Fig.~\ref{fig:SumProduct2}.

The ability to compute the joint distribution in \eqref{eqn:Recursion1} via message passing results in the \ac{map} detector in \eqref{eqn:MAP},  an operation whose burden typically grows exponentially with the block size, with complexity that only grows linearly with $\Blklen$. This is achieved by noting that the \ac{map} estimate satisfies 
\begin{align}
\hat{\myVec{s}}_i\left( \myVec{y}^\Blklen\right)  
=\mathop{\arg \max}\limits_{\myVec{s} \in \mySet{S}} \sum_{  \myStateR_{i\! - \!1}\in \mySet{S}^{\Mem}} &\FwdMsg{f_{i\! - \!1}}{\myState_{i\! - \!1}}(\myStateR_{i\! - \!1}) f(\myVec{y}_{i},  [\myVec{s}_{i\! - \!\Mem\!+\!1}, \ldots, \myVec{s}_{i} ],\myStateR_{i\! - \!1}) \notag \\
& \times  
\BwdMsg{f_{i+1}}{\myStateR_{i}}([\myVec{s}_{i\! - \!\Mem\!+\!1}, \ldots, \myVec{s}_{i} ]), 
\label{eqn:MAP2}
\end{align}
for each $i \in \Blkset$, where the summands can be computed recursively. When the block size $\Blklen$ is large, the messages may tend to zero, and are thus commonly scaled \cite{loeliger2004introduction}, e.g., $\BwdMsg{f_{i+1}}{\myState_i}(\myStateR)$ is replaced with $\gamma_i \BwdMsg{f_{i+1}}{\myState_i}(\myStateR)$ for some scale factor that does not depend on $\myStateR$, and thus does not affect the \ac{map} rule. The BCJR algorithm is summarized as Algorithm~\ref{alg:Algo0}. 

\begin{algorithm}  
	\caption{ The BCJR algorithm}
	\label{alg:Algo0}
	\KwData{Fix an initial forward message $\FwdMsg{f_{0}}{\myState_{0}}(\myStateR)= 1$ and a final backward message $\BwdMsg{f_{0}}{\myState_{\Blklen}}(\myStateR)\equiv 1$. } 
	\For{$i=\Blklen-1,\Blklen-2,\ldots,1$}{
		For each  $\myStateR \in \mySet{S}^\Mem$, compute backward message $ \BwdMsg{f_{i}}{\myState_{i}}(\myStateR)$  via \eqref{eqn:Recursion1Backwards}
	}
	\For{$i=1,2,\ldots,\Blklen$}{
		For each  $\myStateR\in \mySet{S}^\Mem$, compute forward message $ \FwdMsg{f_{i}}{\myState_{i}}(\myStateR)$  via \eqref{eqn:Recursion1Forwards}
	}
	\KwOut{$\hat{\myVec{s}}^\Blklen = [\hat{\myVec{s}}_1, \ldots, \hat{\myVec{s}}_\Blklen]^T$, each obtained using \eqref{eqn:MAP2} 
	}
\end{algorithm}

\paragraph{BCJRNet}
BCJRNet is a receiver method that learns to implement \ac{map} detection  from labeled data. 
BCJRNet exploits the fact that in order to implement Algorithm~\ref{alg:Algo0}, one must be able to specify the factor graph representing the underlying distribution. In particular, the stationarity assumption implies that the complete factor graph is  encapsulated in the single function $f(\cdot)$ \eqref{eqn:FSC_funcNode} {\em regardless of the block size $\Blklen$}. Building upon this insight, BCJRNet utilizes \acp{dnn} to learn the mapping carried out at the function node separately from the inference task. 
\begin{figure}
	\centering
	{\includefig{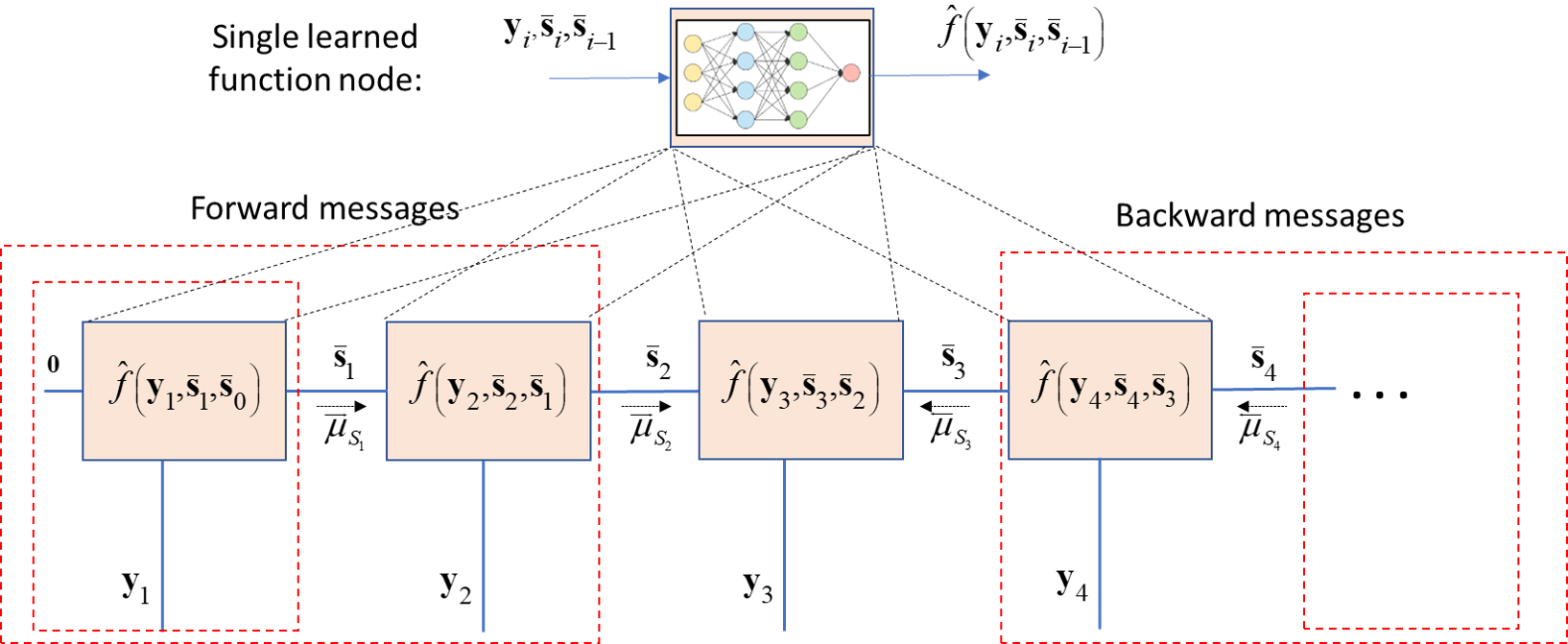}} 
	\caption{An illustration of BCJRNet with a learned stationary factor graph.}
	\label{fig:LeanredSumProduct1}	  
\end{figure}
The resulting learned stationary factor graph is then used to recover  $\{\myS_i\}$ by message passing, as illustrated in Fig. \ref{fig:LeanredSumProduct1}. 
As learning a single function node is expected to be a simpler task compared to learning the overall inference method for recovering $\myVec{S}^\Blklen$ from $\myVec{Y}^\Blklen$, this approach allows using relatively compact \acp{dnn}, which in turn can be learned  from a relatively small set of labeled data. Furthermore, the learned function node describes the factor graph for different values of $\Blklen$.  
When the learned function node is an accurate estimate of the true one, BCJRNet effectively implements the \ac{map} detection rule \eqref{eqn:MAP}, and thus approaches the minimal probability of error. 

The function node that encapsulates the factor graph of stationary finite-memory channels is given in \eqref{eqn:FSC_funcNode}. The formulation in \eqref{eqn:FSC_funcNode} implies that it can be estimated by training an \ac{ml}-based system to evaluate  $\Pdf{\myY_i | \myState_i}(\cdot)$ from which the corresponding function node value is obtained via \eqref{eqn:FSC_funcNode}. 
Once the factor graph representing the channel is learned, symbol recovery is carried out using Algorithm~\ref{alg:Algo0}.  
As the mapping of Algorithm~\ref{alg:Algo0} is invariant to scaling $f(\myVec{y}_i, \myStateR_{i}, \myStateR_{i-1})$ with some factor that does not depend on the states, it follows that a parametric estimate of the function $f(\cdot)$, denoted $\hat{f}_{\myVec{\theta}}(\cdot)$, can be obtained using the same networks utilized for learning the log-likelihood in ViterbiNet. Specifically, the learned log-likelihood is used in \eqref{eqn:FSC_funcNode} to obtain $\hat{f}_{\myVec{\theta}}(\cdot)$.
The resulting receiver, referred to as {\em BCJRNet}, thus implements BCJR detection in a data-driven manner. 

\paragraph{BCJRNet vs. ViterbiNet}
The same \ac{dnn} architecture can  be applied, once trained, to carry out multiple inference algorithms in a hybrid model-based/data-driven manner, including the BCJR scheme (as BCJRNet) as well as the Viterbi algorithm (via ViterbiNet).  Since both BCJRNet and ViterbiNet utilize the same learned models, one can decide which inference system to apply, i.e.,  BCJRNet or ViterbiNet, by considering the differences in the algorithms from which they are derived, i.e., the BCJR algorithm and the Viterbi algorithm, respectively.
The main advantages of Algorithm~\ref{alg:Algo1} over  Algorithm~\ref{alg:Algo0}, and thus of ViterbiNet over BCJRNet, are its reduced complexity and real-time operation. 
In particular, both algorithms implement recursive computations, involving $|\mySet{S}|^{\Mem}$ evaluations for each sample, and thus the complexity of both algorithms grows linearly with the block size $\Blklen$. Nonetheless, the Viterbi scheme computes only a forward recursion and can thus provide its estimations in real time within a given delay from each incoming observation, while the BCJR scheme implements both forward and backward recursions, and can thus infer only once the complete block is observed, while involving twice the computations carried out by the Viterbi detector.

The main advantage of Algorithm~\ref{alg:Algo0} over  Algorithm~\ref{alg:Algo1}, i.e., of BCJRNet over ViterbiNet, stems from the fact that it implements the \ac{map} rule \eqref{eqn:MAP}, which minimizes the error probability. 
The Viterbi algorithm is designed to compute the maximum likelihood {\em sequence} detector, i.e., \begin{equation}
\label{eqn:MLSeq1}
    \hat{\myVec{s}}^{ \Blklen}\left( \myVec{y}^{ \Blklen}\right)  
= \mathop{\arg \max}_{\myVec{s}^{ \Blklen} \in \mySet{S}^\Blklen } \Pdf{\myVec{Y}^{ \Blklen} | \myVec{S}^{ \Blklen}}\left( {\myVec{y}^{ \Blklen} | \myVec{s}^{ \Blklen}}\right),
\end{equation}
which is not equivalent to the symbol-level \ac{map} rule \eqref{eqn:MAP}. To see this, we   focus on the case where the symbols are equiprobable, as in such scenarios the sequence-wise maximum likelihood rule coincides with the  sequence-wise \ac{map} detector. Here, the decision rule implemented by the Viterbi algorithm \eqref{eqn:MLSeq1} can be written as 
\begin{equation}
\label{eqn:MLSeq2}
    \hat{\myVec{s}}^{ \Blklen}\left( \myVec{y}^{ \Blklen}\right)  
= \mathop{\arg \max}_{\myVec{s}^{ \Blklen} \in \mySet{S}^\Blklen } \Pdf{\myVec{S}^{ \Blklen} | \myVec{Y}^{ \Blklen}}\left( {\myVec{s}^{ \Blklen} | \myVec{y}^{ \Blklen}}\right).
\end{equation}
For a given realization $\myVec{Y}^{ \Blklen}=  \myVec{y}^{ \Blklen}$, the function $\Pdf{\myVec{S}^{ \Blklen} | \myVec{Y}^{ \Blklen}}\left( {\myVec{s}^{ \Blklen} | \myVec{y}^{ \Blklen}}\right)$ maximized by the sequence-wise detector in \eqref{eqn:MLSeq2} is a joint distribution measure. The functions $\{\Pdf{\myVec{S}_i | \myVec{Y}^{ \Blklen}}\left( {\myVec{s}_i | \myVec{y}^{ \Blklen}}\right)\}_{i=1}^{\Blklen}$, which are the individually maximized by the symbol-wise \ac{map} rule computed by the BCJR scheme \eqref{eqn:MAP}, are the marginals of the aforementioned joint distribution. Furthermore, given $\myVec{Y}^{ \Blklen}=  \myVec{y}^{ \Blklen}$ the elements of $\myVec{S}^{ \Blklen}$ are statistically dependent in finite-memory channels. As a a result, the maxima of the joint distribution  $\Pdf{\myVec{S}^{ \Blklen} | \myVec{Y}^{ \Blklen}}$ is not necessarily the individual maximas of each of its marginals, i.e., the elements of the vector $\hat{\myVec{s}}^{ \Blklen}\left( \myVec{y}^{ \Blklen}\right) $ in \eqref{eqn:MLSeq1} are not necessarily the symbol-wise \ac{map} estimates $\{\hat{\myVec{s}}_i\left( \myVec{y}^{ \Blklen}\right) \}_{i=1}^{\Blklen}$ in \eqref{eqn:MAP}. To conclude, the Viterbi algorithm does not implement the symbol-wise \ac{map} even in the presence of equal priors. This explains the difference in their performance, since, unlike the BCJR scheme, the Viterbi algorithm does not minimize the error probability.


\subsubsection{Example: DeepSIC for Flat MIMO Channels}
DeepSIC proposed in \cite{shlezinger2019deepSIC} is a \ac{dnn}-aided hybrid algorithm that is based on the iterative \ac{sic} method  \cite{choi2000iterative} for symbol detection in flat \ac{mimo} channels. However, unlike its model-based counterpart, and alternative deep \ac{mimo} receivers such as DetNet, it is not tailored for linear Gaussian channels of the form \eqref{eqn:Gaussian}. The only assumption required is that the channel is memoryless, i.e., $\Mem = 1$, and thus we drop the time index subscript in this example. 
As in our previous \ac{dnn}-aided examples, we first review iterative \ac{sic}, after which we present its \ac{dnn}-aided implementation. 

\begin{figure}
	\centering
	\includegraphics[width = \columnwidth]{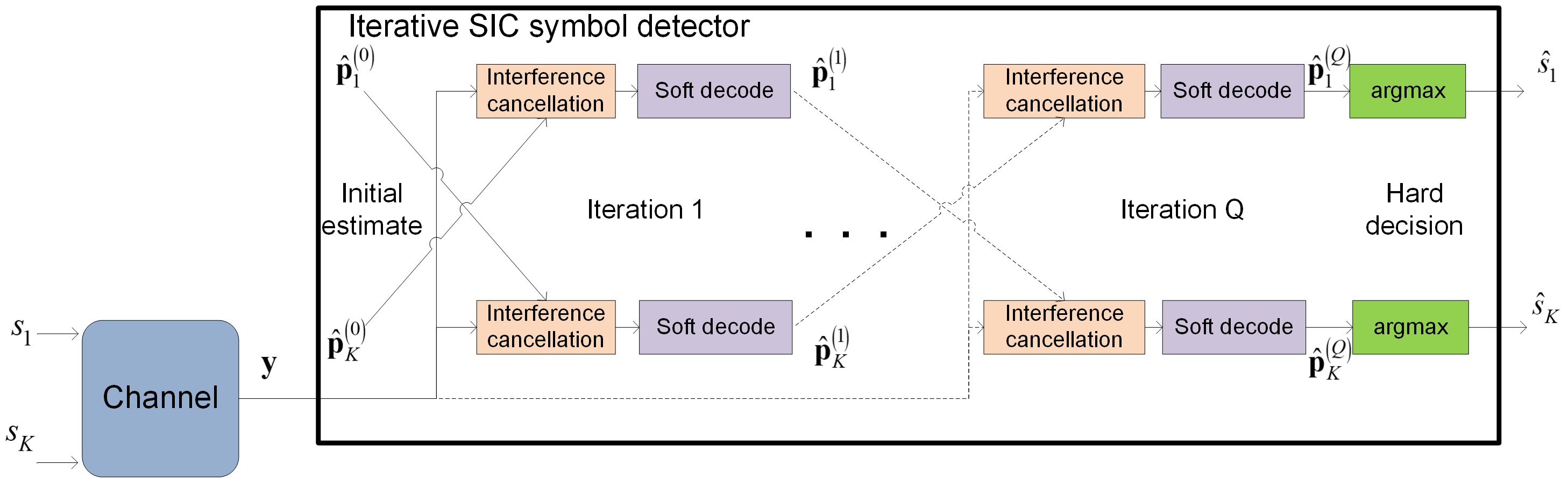}
	\vspace{-0.4cm}
	\caption{Iterative soft interference cancellation illustration.}
	\label{fig:SoftIC1}
\end{figure}

\paragraph{Iterative Soft Interference Cancellation}
The iterative  \ac{sic} algorithm proposed in \cite{choi2000iterative} is a \ac{mimo}  detection method that combines multi-stage interference cancellation with soft decisions.  
The detector operates in an iterative fashion where, in each iteration, an estimate of the conditional \ac{pmf} of $S_k$, which is the $k$th entry of $\myS$, given the observed $\myY = \myVec{y}$, is generated for every symbol $k \in \{1,2,\ldots, \Nusers\} :=\NusersSet$ using the corresponding estimates of the interfering symbols $\{S_l\}_{l \neq k}$ obtained in the previous iteration. Iteratively repeating this procedure refines the conditional distribution estimates, allowing the detector to accurately recover each symbol from the output of the last iteration.  This iterative procedure is illustrated in Fig.~\ref{fig:SoftIC1}.  

To formulate the algorithm, we consider the flat Gaussian \ac{mimo} channel in \eqref{eqn:Gaussian}. 
Iterative \ac{sic} consists of $\Niter$ iterations. Each iteration indexed  $q \in \{1,2,\ldots, \Niter\} \triangleq \NiterSet$ generates $\Nusers$ distribution vectors $\hat{\myVec{p}}_k^{(q)}$ of size $\CnstSize\times 1$, where $k \in \NusersSet$. These vectors are computed from the channel output $\myVec{y}$ as well as the distribution vectors obtained at the previous iteration, $\{ \hat{\myVec{p}}_k^{(q-1)}\}_{k=1}^{\Nusers}$. The entries of  $\hat{\myVec{p}}_k^{(q)}$ are estimates of the distribution of $S_k$ for each possible symbol in $\mathcal{S}$, given the channel output $\myY  = \myVec{y}$ and assuming that the interfering symbols $\{S_l \}_{l \neq k}$ are distributed via $\{ \hat{\myVec{p}}_l^{(q-1)}\}_{l \neq k}$. 
Every iteration consists of two steps, carried out in parallel for each user: {\em Interference cancellation} and {\em soft decoding}. 
Focusing on the $k$th user and the $q$th iteration, the interference cancellation stage first computes the expected values and variances of $\{S_l\}_{l \neq k}$ based on  $\{ \hat{\myVec{p}}_l^{(q-1)}\}_{l \neq k}$. Letting $\{\alpha_j\}_{j=1}^{\CnstSize}$ be the indexed elements of the constellation set $\mySet{S}$, the expected values and variances are computed via
$e_l^{(q-1)} = \sum_{\alpha_j \in \mySet{S}} \alpha_j \big( \hat{\myVec{p}}_l^{(q-1)}\big)_j$, 
and 
$v_l^{(q-1)} = \sum_{\alpha_j \in \mySet{S}}\big(  \alpha_j - e_l^{(q-1)}\big) ^2 \big( \hat{\myVec{p}}_l^{(q-1)}\big)_j$, 
respectively. 
The contribution of the interfering symbols from $\myVec{y}$ is then canceled by replacing them with $\{e_l^{(q-1)}\}$ and subtracting their resulting term. Letting $\myVec{h}_l$ be the $l$th column of $\myMat{H}$, the interference canceled channel output is given by
\begin{align}
\label{eqn:Cancelled1} 
\myVec{Z}_k^{(q)}
&= \myVec{Y}  \!- \!\sum\limits_{l \neq k} \myVec{h}_l e_l^{(q-1)}  
= \myVec{h}_k S_k  \!+\! \sum\limits_{l \neq k} \myVec{h}_l (S_l  \!-\! e_l^{(q-1)}) \!+\! \myVec{W} .
\eqspace
\end{align} 
Substituting the channel output $\myVec{y}$ into \eqref{eqn:Cancelled1}, the realization of the interference canceled $\myVec{Z}_k^{(q)} $, denoted  $\myVec{z}_k^{(q)}$, is obtained. 

To implement soft decoding, it is assumed that $\tilde{W}_k^{(q)}  \triangleq \sum_{l \neq k} \myVec{h}_l (S_l \!- \!e_l^{(q\!-\! 1)}) + \myVec{W} $ obeys a zero-mean Gaussian distribution, independent of $S_k $, and that its covariance is given by
$\CovMat{\tilde{W}_k^{(q)}} = \SigW \myI_{\Nusers} + \sum_{l \neq k}v_l^{(q-1)}  \myVec{h}_l \myVec{h}_l^T$, where $\SigW$ is the noise variance. 
Combining this assumption with \eqref{eqn:Cancelled1}, the conditional distribution of $\myVec{Z}_k^{(q)}$ given $S_k  = \alpha_j$ is multivariate Gaussian with mean value  $\myVec{h}_k \alpha_j$ and covariance $\CovMat{\tilde{W}_k^{(q)}}$. Since $\myVec{Z}_k^{(q)}  $ is given by a bijective transformation of $\myVec{Y} $, it holds that $\Pdf{S_k | \myY}(\alpha_j | \myVec{y}) = \Pdf{S_k |  \myVec{Z}_k^{(q)}}(\alpha_j | \myVec{z}_k^{(q)})$ for each $\alpha_j \in \mySet{S}$ under the above assumptions. Consequently, the conditional distribution of $S_k $ given $\myY $ is approximated from the conditional distribution of $\myVec{Z}_k^{(q)}$ given $S_k $ via Bayes theorem. Since the symbols are equiprobable, this estimated conditional distribution is computed as 
\begin{align}
\left( \hat{\myVec{p}}_k^{(q)}\right)_j %
&= \frac{\exp \left\{-\frac{1}{2} \left( \myVec{z}_k^{(q)} - \myVec{h}_k\alpha_j \right)^T\CovMat{\tilde{W}_k^{(q)}}^{-1} \left( \myVec{z}_k^{(q)} - \myVec{h}_k\alpha_j \right)   \right\} } { \sum\limits_{\alpha_{j'}\in\mySet{S} }\exp \left\{-\frac{1}{2} \left( \myVec{z}_k^{(q)} - \myVec{h}_k\alpha_{j'} \right)^T\CovMat{\tilde{W}_k^{(q)}}^{-1} \left( \myVec{z}_k^{(q)} - \myVec{h}_k\alpha_{j'} \right)   \right\} }.
\label{eqn:CondDist2} 
\end{align}

After the final iteration, the symbols are decoded by taking the symbol that maximizes the estimated conditional distribution for each user, i.e.,  
\begin{equation}
\label{eqn:HardDet}
\hat{s}_k = \mathop{\arg \max}\limits_{j \in \{1,\ldots,\CnstSize\}}\left( \hat{\myVec{p}}_k^{(\Niter)}\right)_j. 
\end{equation}
The overall joint detection scheme is summarized  as Algorithm \ref{alg:Algo1SIC}.
The initial estimates $\{\hat{\myVec{p}}_k^{(0)}\}_{k=1}^{\Nusers}$ can be arbitrarily set. For example, these may be chosen based on a linear separate estimation of each symbol for $\myVec{y}$, as proposed in \cite{choi2000iterative}. 

\begin{algorithm}
	\caption{Iterative Soft Interference Cancellation Algorithm \cite{choi2000iterative}}
	\label{alg:Algo1SIC}
	\KwData { Set $q=1$, and generate an initial guess of  $\{\hat{\myVec{p}}_k^{(0)}\}_{k =1}^{\Nusers}$.} 
	\For{$q=1,2,\ldots,\Niter$}{
		\label{stp:MF1a} Compute the expected values $\{e_k^{(q-1)}\}$ and variances $\{v_k^{q-1}\}$\;
		\label{stp:IC} {\em Interference cancellation:} For each $k \in \NusersSet$ compute $\myVec{z}_k^{(q)}$ via \eqref{eqn:Cancelled1} \;
		\label{stp:SoftDec} {\em Soft decoding:} For each $k \in \NusersSet$, estimate   $\hat{\myVec{p}}_k^{(q)}$ via \eqref{eqn:CondDist2} 
	}	
	\KwOut{ Hard decoded output $\hat{\myVec{s}}$, obtained via \eqref{eqn:HardDet}} 
\end{algorithm} 

Iterative \ac{sic} has several advantages  {compared to both joint decoding as well as separate decoding}: In terms of computational complexity, it replaces the joint exhaustive search over $\mySet{S}^\Nusers$, required by the \ac{map} decoder, with a set of computations carried out separately for each user. Hence, its computational complexity only grows linearly with the number of users \cite{andrews2005interference}, making it feasible with large values of $\Nusers$.  Unlike conventional separate decoding, in which the symbol of each user is recovered individually while treating the interference as noise, the iterative procedure refines the separate estimates sequentially, and the usage of soft values mitigates the effect of error propagation. Algorithm \ref{alg:Algo1SIC} is thus capable of approaching the performance  of the \ac{map} detector, which is only feasible for small values of $\Nusers$.

\paragraph{DeepSIC}
Iterative \ac{sic} is specifically designed for linear channels of the form \eqref{eqn:Gaussian}. In particular, the interference cancellation in Step \ref{stp:IC} of Algorithm \ref{alg:Algo1SIC} requires the contribution of the interfering symbols to be additive. Furthermore, it requires accurate \ac{csi}. To circumvent these limitations in the model-based approach, the \ac{dnn}-aided DeepSIC learns to implement the iterative \ac{sic} from data.

{\bf Architecture:}
DeepSIC builds upon the observation that  iterative \ac{sic}  can be viewed as a set of interconnected basic building blocks, each implementing the two stages of interference cancellation and soft decoding, i.e., Steps \ref{stp:IC}-\ref{stp:SoftDec} of Algorithm \ref{alg:Algo1SIC}. While the high level architecture in Fig. \ref{fig:SoftIC1} is ignorant of the underlying channel model, the basic building blocks are channel-model-dependent. In particular, interference cancellation requires the contribution of the interference to be additive, i.e., a linear model channel as in \eqref{eqn:Gaussian}, as well as full \ac{csi}, in order to cancel the contribution of the interference. Soft decoding requires complete knowledge of the channel input-output relationship in order to estimate the conditional probabilities via \eqref{eqn:CondDist2}. 

\begin{figure}
	\centering
	\includegraphics[width = \columnwidth]{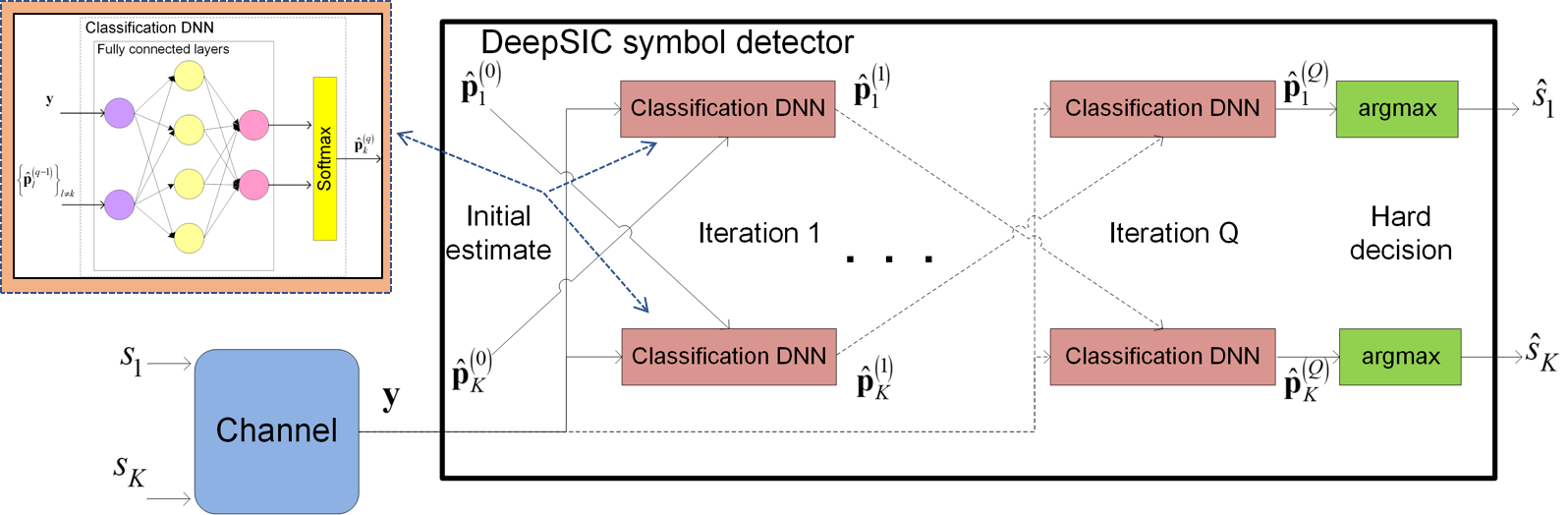} 
	\vspace{-0.4cm}
	\caption{Deep\ac{sic} illustration.}
	\label{fig:DeepSoftIC1}
\end{figure}

Although each of these basic building blocks consists of two sequential procedures that are completely channel-model-based, we note that the purpose of these computations is to carry out a classification task. In particular, the $k$th building block of the $q$th iteration, $k \in \NusersSet$, $q  \in \NiterSet$, produces $\hat{\myVec{p}}_k^{(q)}$, which is an estimate of the conditional \ac{pmf} of $S_k $ given $\myY  = \myVec{y}$ based on $\{\hat{\myVec{p}}_l^{(q-1)}\}_{l\neq k}$. Such computations are naturally implemented by classification \acp{dnn}, e.g., fully-connected networks with a softmax output layer. 
Embedding these \ac{ml}-based conditional \ac{pmf} computations into the iterative \ac{sic} block diagram in Fig. \ref{fig:SoftIC1} yields the overall receiver architecture depicted in Fig. \ref{fig:DeepSoftIC1}. The initial estimates $\{\hat{\myVec{p}}_k^{(0)}\}_{k=1}^{\Nusers}$ can be set to represent a uniform distribution, i.e., $\big(\hat{\myVec{p}}_k^{(0)}\big)_j \equiv \frac{1}{\CnstSize}$. 

A major advantage of using classification \acp{dnn} as the basic building blocks in Fig. \ref{fig:DeepSoftIC1} stems from the fact that such \ac{ml}-based methods are capable of accurately computing conditional distributions in   complex non-linear setups without requiring a-priori knowledge of the channel model and its parameters. Consequently, when these building blocks are trained to properly implement their classification task, the  receiver essentially realizes iterative \ac{sic} for arbitrary channel models in a data-driven fashion.

{\bf Training Methods:} 
In order for the \ac{dnn}-aided receiver structure of Fig.~\ref{fig:DeepSoftIC1} to  reliably implement joint decoding, its building block classification \acp{dnn} must be properly trained. Here, we consider two possible approaches to train the receiver based on a set of $\Ntraining$ pairs of channel inputs and their corresponding outputs, denoted $\{\myVec{s}_t, \myVec{y}_t \}_{t=1}^{\Ntraining}$: {\em End-to-end training}, and {\em sequential training}.

{\em End-to-end training}: The first approach jointly trains the entire network, i.e., all the building block \acp{dnn}. Since the output of the deep network is the set of conditional distributions  $\{\hat{\myVec{p}}_k^{(\Niter)}\}_{k=1}^{\Nusers}$, where each  $\hat{\myVec{p}}_k^{(\Niter)}$ is used to estimate $S_k $, we use the sum cross entropy  as the training objective. Let 
$\myVec{\theta}$ be the network parameters, and 
$\hat{\myVec{p}}_k^{(\Niter)}(\myVec{y}, \alpha; \myVec{\theta} )$ be the entry of $\hat{\myVec{p}}_k^{(\Niter)}$ corresponding to $S_k  = \alpha$ when the input to the network parameterizd by $\myVec{\theta}$ is $\myVec{y}$.
The sum cross entropy loss is 
\begin{equation}
\label{eqn:SumCE}
\mathcal{L}_{\rm SumCE} (\myVec{\theta})= \frac{1}{\Ntraining}\sum_{t=1}^{\Ntraining}
\sum_{k=1}^{\Nusers} -\log \hat{\myVec{p}}_k^{(\Niter)}\big(\myVec{y}_t, (\myVec{s}_t)_k ; \myVec{\theta}\big). 
\end{equation} 

Training the receiver in Fig. \ref{fig:DeepSoftIC1} in an end-to-end manner based on the loss \eqref{eqn:SumCE} jointly updates the coefficients of all the $\Nusers \cdot \Niter$ building block \acp{dnn}. For a large number of users, training so many parameters simultaneously is expected to require a large labeled set. 

{\em Sequential training}: To allow the network to be trained with a reduced number of training samples, we note that the goal of each building block \ac{dnn} does not depend on the iteration index: The $k$th building block of the $q$th iteration outputs a soft estimate of $S_k $ for each $q \in \NiterSet$. Therefore, each building block \ac{dnn} can be trained individually, by minimizing the conventional cross entropy loss. To formulate this objective, let  
$\myVec{\theta}_{k}^{(q)}$ represent the parameters of the $k$th \ac{dnn} at iteration $q$, and write 
$\hat{\myVec{p}}_k^{(q)}\big(\myVec{y}, \{\hat{\myVec{p}}_l^{(q-1)}\}_{l\neq k}, \alpha; \myVec{\theta}_{k}^{(q)}\big)$ as the entry of $\hat{\myVec{p}}_k^{(q)}$ corresponding to $S_k = \alpha$ when the \ac{dnn} parameters are $\myVec{\theta}_{k}^{(q)}$ and its inputs are $\myVec{y}$ and  $\{\hat{\myVec{p}}_l^{(q-1)}\}_{l\neq k}$. The cross entropy loss is given by
\begin{equation}
\label{eqn:CE}
\mathcal{L}_{\rm CE}(\myVec{\theta}_{k}^{(q)})  = \frac{1}{\Ntraining}\sum_{t=1}^{\Ntraining}
-\log \hat{\myVec{p}}_k^{(q)}\big(\tilde{\myVec{y}}_t, \{\hat{\myVec{p}}_{t,l}^{(q-1)}\}_{l\neq k}, (\tilde{\myVec{s}}_t)_k ; \myVec{\theta}_{k}^{(q)}\big),
\end{equation}	
where $\{\hat{\myVec{p}}_{t,l}^{(q-1)}\}$ represent the estimated probabilities associated with $\myVec{y}_i$ computed at the previous iteration.
The problem with training each \ac{dnn} individually is that the soft estimates $\{\hat{\myVec{p}}_{t,l}^{(q-1)}\}$ are not provided as part of the training set. This challenge can be tackled by training the \acp{dnn} corresponding to each layer in a sequential manner, where for each layer the outputs of the trained previous iterations are used as the soft estimates fed as training samples. 

Sequential training uses the $\Ntraining$ input-output pairs to train each \ac{dnn} individually. Compared to the end-to-end training that utilizes the training samples to learn the complete set of parameters, which can be quite large, sequential training uses the same data set  to learn a significantly smaller number of parameters, reduced by a factor of $\Nusers\cdot \Niter$, multiple times. Consequently, this approach is expected to require much fewer training samples, at the cost of a longer learning procedure for a given training set, due to its sequential operation, and possible performance degradation as the building blocks are not jointly trained.

\subsubsection{Summary}
\ac{dnn}-aided algorithms implement hybrid model-based/data-driven inference by integrating \ac{ml} into established model-based methods. As such, it is particularly suitable for digital communications setups, in which a multitude of reliable model-based algorithms exist, each tailored to a specific structure. The implementation of these techniques in a data-driven fashion thus has three main advantages as a model-based \ac{ml} strategy: First, when properly trained, the resulting system effectively implements the model-based algorithm from which it originated, thus benefiting from its proven performance and controllable complexity, while being robust to \ac{csi} uncertainty and operable in complex environments, due to the usage of \acp{dnn}. This behavior is numerically illustrated in the simulation study detailed in Section \ref{subsec:SymbolDetectionSims}. 

Second, the fact that \ac{dnn}-aided algorithms use \ac{ml} tools as intermediate components in the overall end-to-end inference tasks allows the use of relatively compact networks which can be trained with small training sets. Even when the overall system consists of a large set of \acp{dnn}, as is the case in  DeepSIC, their interpretable operation which follows from the model-based method facilitates their training with small data sets, e.g., via sequential training techniques. 

Finally, \ac{dnn}-aided algorithms can utilize different levels of domain knowledge, depending on what prior information one has on the problem at hand. For example, BCJRNet requires only prior knowledge that the channel has finite memory to learn to carry out \ac{map} detection from data. When additional domain knowledge is available, such as an underlying stationarity or some partial \ac{csi}, it can be incorporated into the number and structure of the learned function nodes, further reducing the number of training data required to tune the receiver. The resulting ability of \ac{dnn}-aided symbol detectors to adapt with small training sets  can be exploited to facilitate channel tracking via periodic re-training using existing pilots and other forms of structures present in digital communications protocols, as demonstrated in \cite{shlezinger2019viterbinet,shlezinger2019deepSIC}.

\subsection{Numerical Study}
\label{subsec:SymbolDetectionSims}
In this section, we present a numerical study of the aforementioned symbol detection mechanisms. We begin with considering finite-memory channels, for which we evaluate the data-driven \ac{sbrnn} receiver, ViterbiNet, and BCJRNet, comparing them to the model-based detection methods for such channels. Then we consider memoryless \ac{mimo} channels, where we compare the data-driven DetNet and DeepSIC to model-based detection. 

\subsubsection{Finite-Memory Channel}
We first numerically evaluate the performance of the \ac{dnn}-aided ViterbiNet and BCJRNet, and compare this performance to that of the conventional model-based Viterbi algorithm and BCJR detector, as well as to that of the \ac{sbrnn} receiver detailed in Section \ref{subsec:SymbolDetectionEstablished}. 
Both ViterbiNet and BCJRNet are implemented using the classification architecture in Fig. \ref{fig:LearnedFunctionNode}(a)  with three fully-connected layers: a $1 \times 100$ layer followed by a $100 \times 50$ layer and a $50 \times  16 (=|\CnstSize|^{\Mem})$ layer, using intermediate sigmoid and ReLU activation functions, respectively. 
For the \ac{sbrnn} receiver, we use BRNN length of $\BRNNLen = 10$ with 3 layers of LSTM cell blocks of size 100, and a dropout rate of 0.1.   The networks are trained using $5000$ training samples, which is of the same order and even smaller compared to typical preamble sequences in wireless communication systems.  

We consider two finite-memory  channels: An  \ac{awgn} channel and a Poisson channel, both with memory length of $\Mem = 4$. 
For the  \ac{awgn} channel, we let $W[i]$ be a zero-mean unit variance \ac{awgn} independent of $S[i]$, and let $\myVec{h} (\gamma)\in \mySet{R}^\Mem$ be the channel vector obeying an exponentially decaying profile  $\left( \myVec{h}\right)_\tau \triangleq e^{-\gamma(\tau-1)}$ for $\gamma > 0$. The  input-output relationship is given by
\begin{equation}
\label{eqn:AWGNCh1}
Y[i] = \sqrt{\rho} \cdot\sum\limits_{\tau=1}^{\Mem} \left( \myVec{h}(\gamma)\right)_\tau S[i-\tau + 1] + W[i],
\end{equation}
where $\rho > 0$ represents the \acs{snr}.  
The channel input is randomized from a \ac{bpsk} constellation, i.e., $\mySet{S} = \{-1, 1\}$. 
For the Poisson channel, the symbols represent on-off keying, namely, $\mySet{S} = \{0,1\}$, and the channel output $Y[i]$ is generated  via 
\begin{equation}
\label{eqn:PoissonCh1}
Y[i] | \myVec{S}^\Blklen\sim \mathbb{P}\left( \sqrt{\rho} \cdot\sum\limits_{\tau=1}^{\Mem} \left( \myVec{h}(\gamma)\right)_\tau S[i-\tau + 1] + 1\right),
\end{equation}
where $\mathbb{P}(\lambda)$ is the Poisson distribution with parameter $\lambda > 0$, and $X\sim f(X)$ indicates that the random variable $X$ is distributed according to $f(X)$.

For each channel, we numerically compute the \ac{ser}  for  different values of the \acs{snr} parameter $\rho$. 
For each \acs{snr} $\rho$, the \ac{ser} values are averaged over $20$ different channel vectors  $\myVec{h} (\gamma)$, obtained by letting $\gamma$ vary in the range $[0.1, 2]$. 
For comparison, we  numerically compute the \ac{ser} of the Viterbi and BCJR algorithms. 
\label{txt:Robustness1}
In order to study the resiliency of the data-driven detectors to inaccurate training, we also compute the performance when the receiver only has access to a noisy estimate of $\myVec{h}(\gamma)$, and specifically, to a copy of $\myVec{h}(\gamma)$ whose entries are corrupted by i.i.d. zero-mean Gaussian noise with variance $\sigma_e^2$. In particular, we use  $\sigma_e^2 = 0.1$ for the Gaussian channel \eqref{eqn:AWGNCh1}, and $\sigma_e^2 = 0.08$ for the Poisson channel \eqref{eqn:PoissonCh1}. 
We consider two cases: {\em Perfect \ac{csi}}, in which the channel-model-based detectors have accurate knowledge of  $\myVec{h}(\gamma)$, while the data-driven receivers are trained using labeled samples generated with the same  $\myVec{h}(\gamma)$ used for generating the test data; and {\em \ac{csi} uncertainty}, where the model-based algorithms are implemented with the log-likelihoods (for Viterbi algorithm) and function nodes (for BCJR detection) computed using the noisy version of  $\myVec{h}(\gamma)$,  while the training data is generated with  the noisy version of $\myVec{h}(\gamma)$ instead of the true one. 
In all cases, the information symbols are uniformly randomized in an i.i.d. fashion from $\mySet{S}$, and the test samples are generated from their corresponding channel
with the true  channel vector $\myVec{h}(\gamma)$. 

\begin{figure}
    \centering
    \includegraphics[width=0.7\columnwidth]{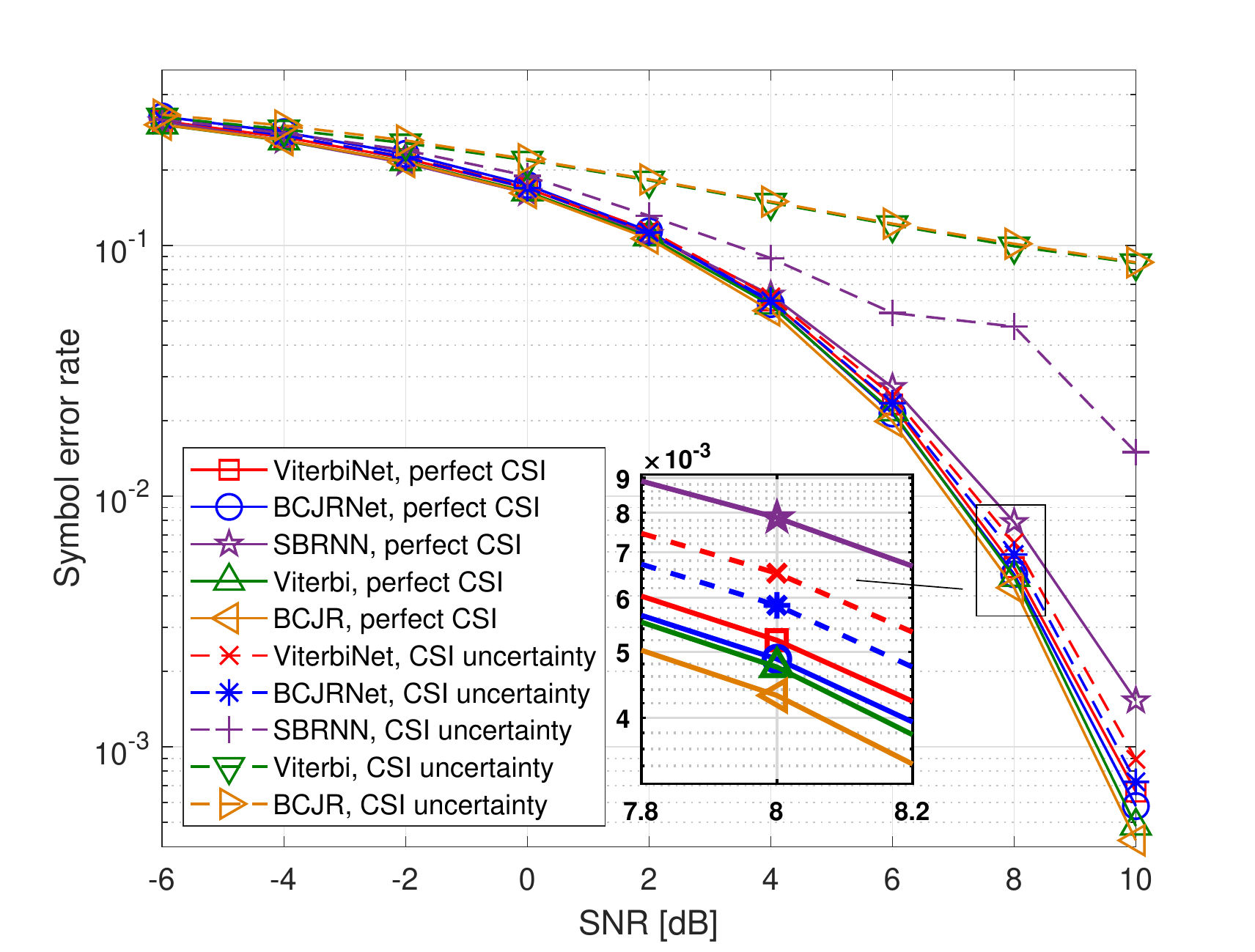}
    \caption{Symbol error rate of different receiver structures for the  \ac{awgn} channel with exponentially decaying taps.}
    \label{fig:AWGN}
\end{figure}

\begin{figure}
    \centering
    \includegraphics[width=0.7\columnwidth]{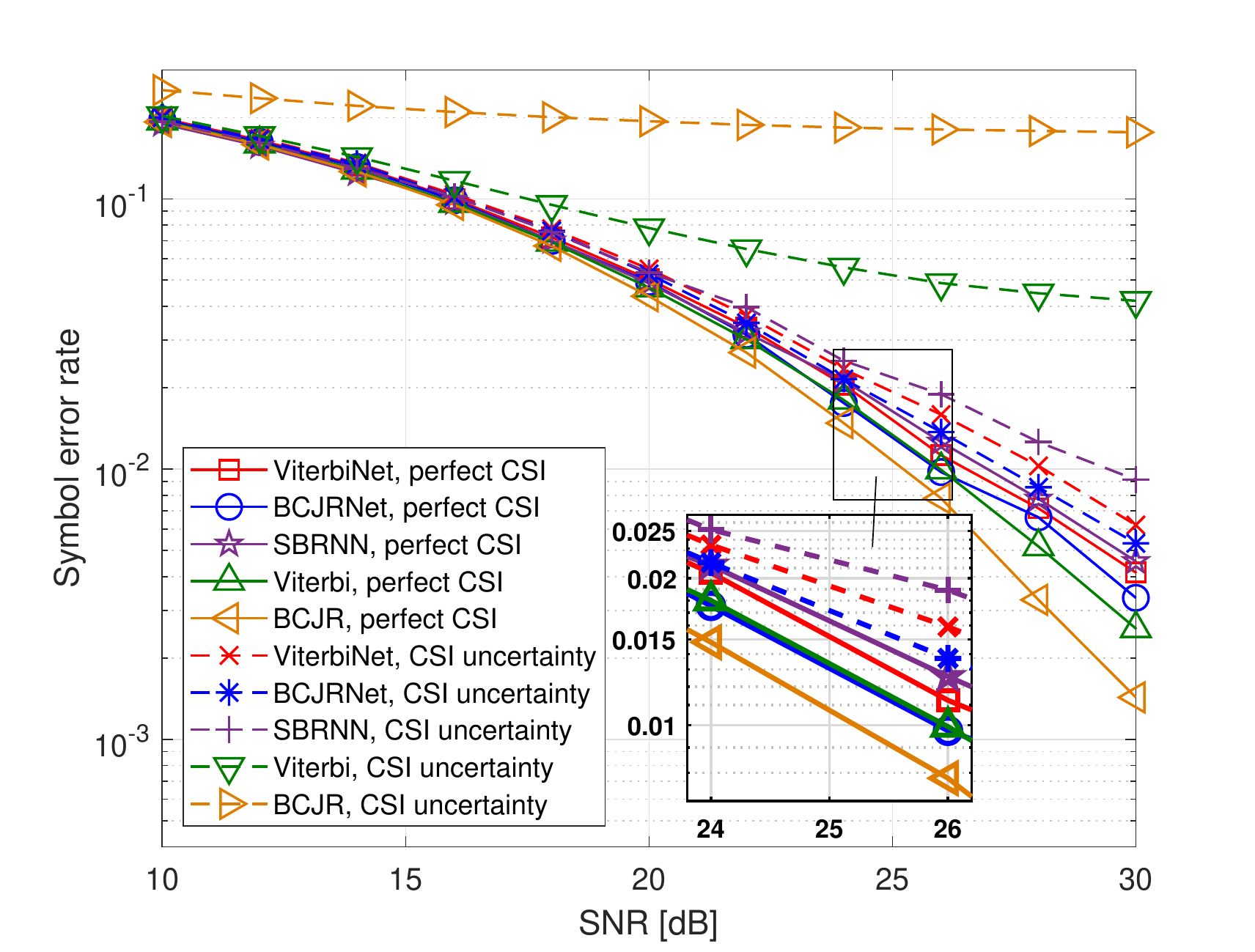}
    \caption{Symbol error rate of different receiver structures for the  Poisson channel.}
    \label{fig:Poisson}
\end{figure}

The numerically computed \ac{ser} values, averaged over $50000$ Monte Carlo simulations, versus $\rho \in [-6,10]$ dB for the \ac{awgn} channel are depicted in Fig. \ref{fig:AWGN}, while the corresponding performance versus $\rho \in [10,30]$ dB for the Poisson channel are depicted in Fig. \ref{fig:Poisson}. Observing Figs. \ref{fig:AWGN}-\ref{fig:Poisson}, we note that the performance of the data-driven receivers approaches that of their corresponding \ac{csi}-based counterparts. 
We also observe that the \ac{sbrnn} receiver, which was shown in \cite{farsad2018neural} to approach the performance of the \ac{csi}-based Viterbi algorithm when sufficient training is provided, is outperformed by ViterbiNet and BCJRNet here due to the small training set size. 
These results demonstrate that our \ac{dnn}-aided detectors, which use compact \ac{dnn} structures embedded into model-based algorithms, require significantly less training compared to symbol detectors based on using established \acp{dnn} for end-to-end inference.

In the presence of \ac{csi} uncertainty, it is observed in Figs. \ref{fig:AWGN}-\ref{fig:Poisson} that both ViterbiNet and BCJRNet significantly outperform the model-based algorithms from which they originate. In particular, when ViterbiNet and BCJRNet are trained with a variety of different channel conditions, they are still capable of achieving relatively good \ac{ser} performance under each of the channel conditions for which they are trained, while the performance of the conventional Viterbi and BCJR algorithms is significantly degraded in the presence of imperfect \ac{csi}. While the \ac{sbrnn} receiver is shown to be more resilient to inaccurate \ac{csi} compared to the Viterbi and BCJR algorithms, it is outperformed by  ViterbiNet  and BCJRNet with the same level of uncertainty, and the performance gap is more notable in the \ac{awgn} channel. 

\begin{figure}
    \centering
    \includegraphics[width=0.7\columnwidth]{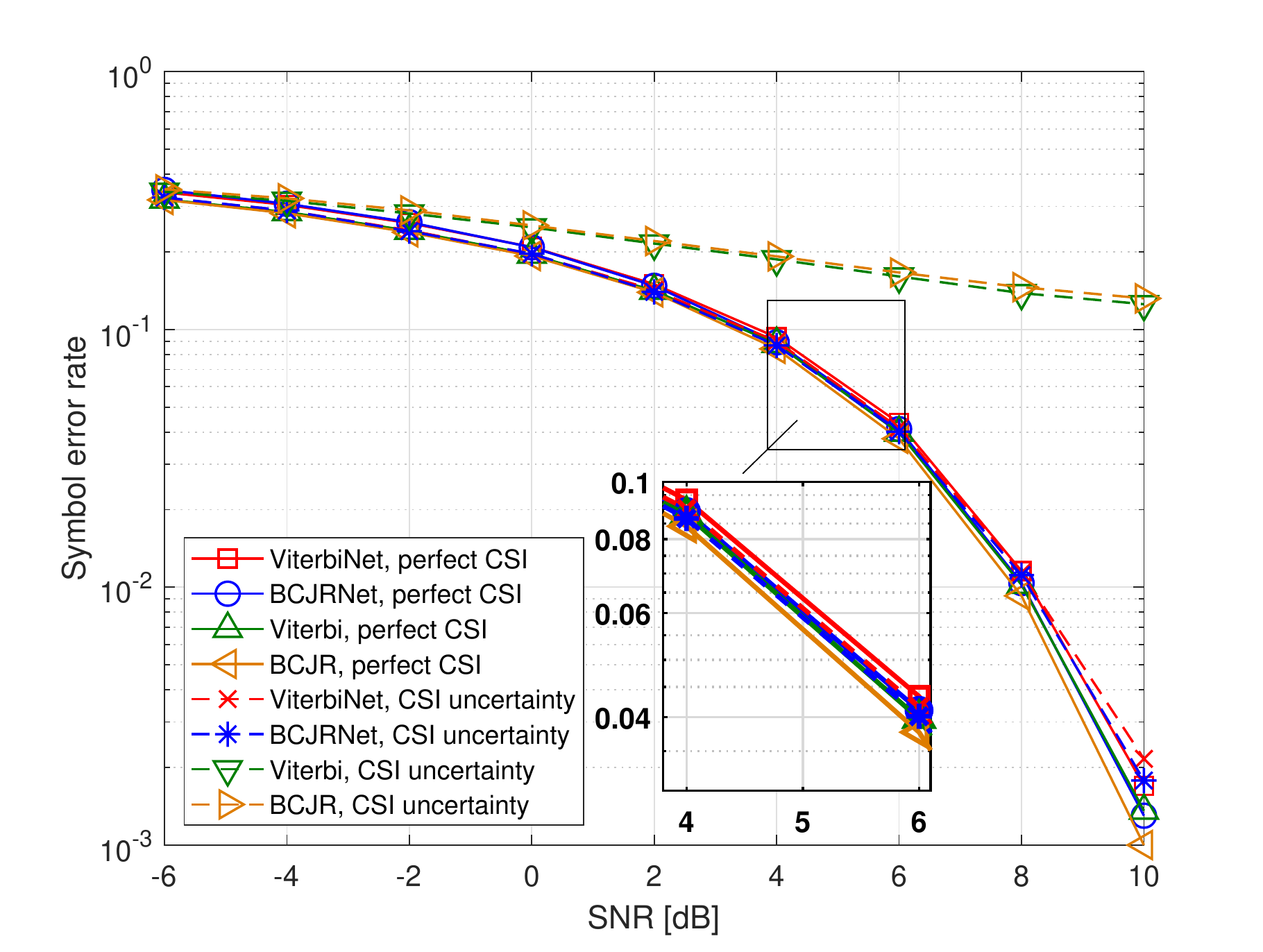}
    \caption{Symbol error rate of different receiver structures for the \ac{awgn} channel with taps generated from the COST2100 model.}
    \label{fig:COST}
\end{figure}

Finally, we evaluate the application of ViterbiNet and BCJRNet for practical channel models. To that aim, we generate 10 realizations from the established COST2100 model \cite{liu2012cost}, which is a widely used model for current cellular communication channels. In particular, we  use the semi-urban 300MHz line-of-sight configuration evaluated in \cite{zhu2013cost} with a single antenna element. The channel output is corrupted by \ac{awgn}, and the symbol detectors operate assuming that the channel has $\Mem=4$ taps. The remaining simulation parameters are the same as those used in Fig.~\ref{fig:AWGN}. The results, depicted in Fig.~\ref{fig:COST}, demonstrate that the ability of ViterbiNet and BCJRNet to approach their model-based counterparts with perfect \ac{csi}, as well as achieve improved performance in the presence of \ac{csi} uncertainty, holds for practical channel models. 

\subsubsection{Memoryless MIMO Channel}
Next, we numerically compare DeepSIC and DetNet for symbol detection in memoryless \ac{mimo} channels. 
In the implementation of the \ac{dnn}-based building blocks of Deep\ac{sic},  we used a different fully-connected network for each training method: For end-to-end training, where all the building blocks are jointly trained, we used a compact network consisting of a  $(\Nantennas + \Nusers - 1) \times 60$ layer followed by ReLU activation and a $60 \times \CnstSize$  layer. 
For sequential training, which sequentially adapts subsets of the building blocks and can thus tune  more parameters using the same training set  (or, alternatively,  requires a smaller training set) compared to end-to-end training, we used three fully-connected layers:  An $(\Nantennas + \Nusers - 1) \times 100$ first layer, a $100 \times 50$ second layer, and a $50 \times \CnstSize$ third layer, with a sigmoid and a ReLU intermediate activation functions, respectively.  
In both iterative \ac{sic} as well as Deep\ac{sic}, we set the number of iterations to $\Niter = 5$. 
Following \cite{samuel2019learning}, DetNet is implemented with $\Niter=90$ layers with a hidden sub-layer size of $8\Nusers$. The data-driven receivers are trained with a relatively small data set of $5000$ training samples, and tested over $20000$ symbols. 

We first consider a linear \ac{awgn} channel as in \eqref{eqn:Gaussian} with a relatively small $\Nusers$. Recall that iterative \ac{sic} as well as DetNet are all designed for such channels. Consequently,  the following study compares the performance of Deep\ac{sic} and DetNet to that of the model-based iterative \ac{sic} as well as the \ac{map} rule \eqref{eqn:MAP} in a scenario for which all schemes are applicable. 
%
The model-based \ac{map} and iterative \ac{sic} detectors, as well as DetNet, all require \ac{csi}, and specifically, accurate knowledge of the channel matrix $\myMat{H}$.  Deep\ac{sic} operates without a-priori knowledge of the channel model and its parameters, learning the decoding mapping from a training set sampled from the considered input-output relationship. In order to compare the robustness of the  detectors to \ac{csi} uncertainty, we also evaluate them when the receiver has access to an estimate of $\myMat{H}$ with entries corrupted by i.i.d. additive Gaussian noise whose variance is given by $\SigE$ times the magnitude of the corresponding entry, where $\SigE > 0$ is referred to as the {\em error variance}. For Deep\ac{sic}, which is model-invariant, we compute the \ac{ser} under \ac{csi} uncertainty by using a training set whose samples are randomized from a channel in which the true $\myMat{H}$ is replaced with its noisy version. 

\begin{figure}
    \centering
    \includegraphics[width=0.7\columnwidth]{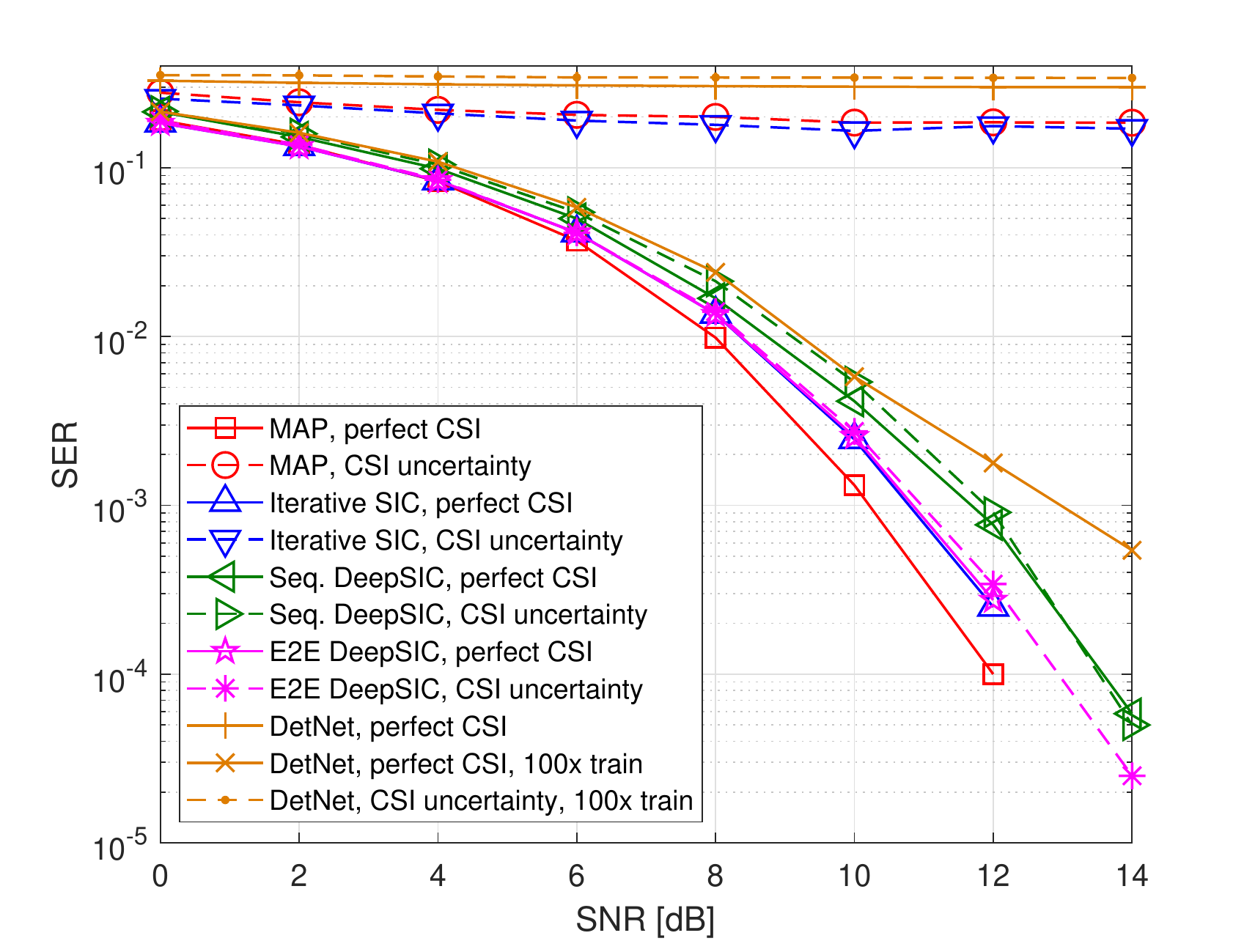}
    \caption{Symbol error rate of different receiver structures for the $6\times6$ \ac{awgn} channel.}
    \label{fig:AWGN6}
\end{figure}

We simulate the $6 \times 6$ linear Gaussian channel, i.e., $\Nusers = 6$ users and $\Nantennas = 6$ receive antennas. The symbols are randomized from a \ac{bpsk} constellation, namely, $\mySet{S} = \{-1, 1\}$ and $\CnstSize = |\mySet{S}| = 2$. The channel matrix $\myMat{H}$ models spatial exponential decay, and its entries are given by
$\left( \myMat{H}\right)_{i,k} = e^{-|i-j|}$, for each $i \in \{1,\ldots, \Nantennas\}$, $ k \in \NusersSet$.  
For each channel, the \ac{ser} of the  receivers is evaluated for both perfect \ac{csi}, i.e., $\SigE = 0$, as well as \ac{csi} uncertainty, for which we use $\SigE = 0.1$.  The  evaluated \ac{ser}  versus the \acs{snr}, defined as $1/\SigW$, is depicted in Fig. \ref{fig:AWGN6}. 

Observing Fig. \ref{fig:AWGN6}, we note that the performance of Deep\ac{sic} with end-to-end training approaches that of the model-based iterative \ac{sic} algorithm, which is within a small gap of the  \ac{map} performance. This demonstrates the ability of Deep\ac{sic} to implement iterative \ac{sic} in a data-driven fashion. The sequential training method, whose purpose is to allow Deep\ac{sic} to train with smaller data sets compared to end-to-end training, also achieves  \ac{ser}  values comparable to iterative \ac{sic}. DetNet, which trains a large number of parameters in an end-to-end fashion, requires $100$ times more training to approach such performance. In the presence of \ac{csi} uncertainty, Deep\ac{sic} is observed to substantially outperform the model-based iterative \ac{sic} and \ac{map} receivers, as well as DetNet operating with a noisy version of $\myMat{H}$ and trained with a hundred times more samples. In particular, it follows from Fig. \ref{fig:AWGN6} that a relatively minor  error of variance   $\SigE = 0.1$ severely deteriorates the performance of the model-based methods, while   Deep\ac{sic} is hardly affected by the same level of \ac{csi} uncertainty. 

\begin{figure}
    \centering
    \includegraphics[width=0.7\columnwidth]{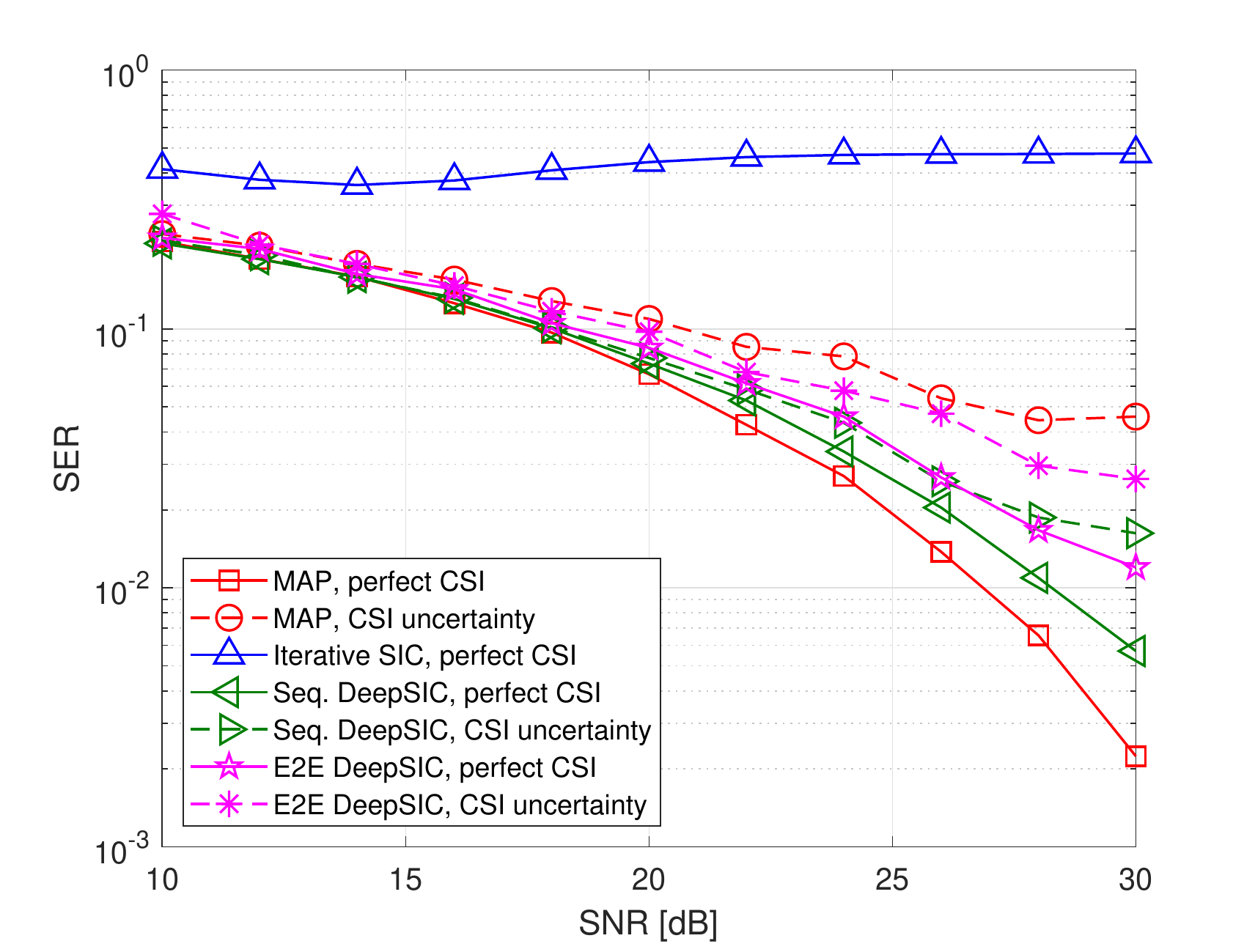}
    \caption{Symbol error rate of different receiver structures for the $4\times 4$ Poisson channel.}
    \label{fig:Poisson4}
\end{figure}


Next, we consider a Poisson channel. We use $\Nusers = 4$ and $\Nantennas = 4$. Here, the symbols are randomized from an on-off keying for which $\mySet{S} = \{0,1\}$. The entries of the channel output are related to the input via the conditional distribution 
\begin{equation}
\label{eqn:PoissonChannel}
\left( \myY[i]\right)_j  | \myS[i]  \sim \mathbb{P}\left( \frac{1}{\sqrt{\SigW}}\left( \myMat{H} \myS[i]\right)_j  + 1\right), \qquad j \in \{1,\ldots, \Nantennas\}. 
\end{equation}
As DetNet is designed for linear Gaussian channels, DeepSIC is the only data-driven receiver evaluated for this channel.

The achievable \ac{ser} of Deep\ac{sic} versus \acs{snr} under both perfect \ac{csi} as well as \ac{csi} uncertainty with error variance $\SigE = 0.1$ is compared to the \ac{map} and iterative \ac{sic} detectors in Fig. \ref{fig:Poisson4}. Observing Fig. \ref{fig:Poisson4}, we again note that the performance of Deep\ac{sic} is only within a small gap of the \ac{map} performance with perfect \ac{csi}, and that the data-driven receiver is more robust to \ac{csi} uncertainty compared to the model-based \ac{map}. 
In particular, Deep\ac{sic} with sequential training, which utilizes a deeper network architecture for each building block, outperforms here end-to-end training with basic two-layer structures for the conditional distribution estimation components. We conclude that under such non-Gaussian channels, more complex \ac{dnn} models are required to learn to cancel interference and carry out soft detection accurately. 
Furthermore,  iterative \ac{sic}, which is designed for linear Gaussian channels \eqref{eqn:Gaussian} where interference is additive, achieves very poor performance when the channel model is substantially different from \eqref{eqn:Gaussian}. 
These results demonstrate the ability of Deep\ac{sic} to achieve excellent performance through learning from data for statistical models where model-based interference cancellation is effectively inapplicable.

%
%
%
%
%
%
%

\section{Summary}
\label{sec:Summary}
Deep learning brings forth capabilities that can substantially contribute to future communications systems in tackling some of their expected challenges.
{In particular digital communications systems can significantly benefit from properly harnessing the power of deep learning and its model-agnostic nature. A successful integration of deep learning into communication devices can thus pave the way to reliable and robust communications in various setups, including environments where accurate statistical channel models are scarce or costly to obtain.}
However, digital communications setups are fundamentally different from applications in which deep learning has been extremely successful to date, such as computer vision. In particular,  digital communication exhibit an extremely large number of possible outputs, as these outputs grow exponentially with the modulation order and the block length. 
They also have channel conditions that vary dynamically, and require low computation complexity when used on small battery-powered devices.   
Consequently, in order to achieve the potential benefits of \ac{dnn}-aided communications, researchers and system designers must go beyond the straight-forward application of \acp{dnn} designed for computer vision and natural language processing. A candidate strategy to utilize \acp{dnn} while accounting for the unique characteristics of digital communications setups, as well as the established knowledge of model-based communication methods accumulated over the last decades, is based on model-based \ac{ml}, as detailed in this chapter. 

We reviewed two main strategies for combining data-driven deep learning with model-based methods for digital communications, as well as discussing the need for such hybrid schemes due to the shortcomings of the extreme cases of purely data-driven and solely model-based methods. For each strategy, we presented the main steps in the design of the data-driven systems, and provided concrete examples, all in the context of the basic communication task of symbol detection. We first discussed how established \ac{dnn} architectures can be utilized as symbol detectors, presenting the \ac{sbrnn} receiver of \cite{farsad2018neural} as an example. Then we detailed how the framework of deep unfolding, which designs \acp{dnn} based on iterative optimization algorithms, can give rise to hybrid model-based/data-driven receivers, presenting DetNet of \cite{samuel2019learning} as an example. We identified that the main drawback of these aforementioned techniques in the context of digital communications stems from their usage of highly-parameterized \acp{dnn} applied in an end-to-end fashion, which directly results in the need for massive data sets for training. Then, we presented \ac{dnn}-aided algorithms, where \acp{dnn} are integrated into existing model-based algorithms. We identified the latter as being extremely suitable for digital communications due to the wide variety of model-based algorithms designed for such setups, combined with its ability to incorporate different levels of domain knowledge as well as utilize compact \acp{dnn} as intermediate components in the inference process. 

The \ac{dnn}-aided symbol detectors presented as examples here, i.e., ViterbiNet \cite{shlezinger2019viterbinet}, BCJRNet \cite{shlezinger2020data}, and DeepSIC \cite{shlezinger2019deepSIC}, all numerically demonstrated improved performance over competing strategies when a limited amount of training data is available. {In particular, it is demonstrated that these \ac{dnn}-aided symbol detectors, which are designed to operate in a hybrid model-based/data-driven fashion, learn to approach the performance achieved by purely model-based approaches operating with perfect knowledge of the underlying channel model and its parameters. Furthermore, the \ac{dnn}-aided symbol detectors were shown to be notably more resilient to CSI uncertainty compared to model-based schemes,  carrying out accurate detection in the presence of inaccurate CSI.  Finally, it was demonstrated that model-based deep learning enables \ac{dnn}-aided receivers to learn their mapping from relatively small data sets, making}
it an attractive approach to combine with tracking of dynamic channel conditions.

\bibliographystyle{IEEEtran}
\bibliography{IEEEfull,ModelBasedRefs}


  \backmatter



  \cleardoublepage





\end{document}